\documentclass{vldb}
\makeatletter
\let\old@citex\@citex
\makeatother
\usepackage[english]{babel}
\makeatletter
\let\@citex\old@citex
\makeatother
\usepackage{mathptmx,helvet}

\usepackage{graphicx}
\usepackage{balance}
\usepackage{times}
\usepackage{subfigure}
\usepackage{url}
\usepackage{comment}

\begin{document}

\title{Quegel: A General-Purpose Query-Centric Framework for Querying Big Graphs}

\author{
{Da Yan$^{*1}$,\ \ \ \ James Cheng$^{*2}$,\ \ \ \ M.\ Tamer \"{O}zsu$^{\dag3}$,\ \ \ \ Fan Yang$^{*4}$,}\\
{\ \ \ \ Yi Lu$^{*5}$,\ \ \ \ \ \ John C.\ S.\ Lui$^{*6}$,\ \ \ \ Qizhen Zhang$^{*7}$,\ \ \ \ \ Wilfred Ng$^{+8}$}
\vspace{1.6mm}\\
\fontsize{10}{10}\selectfont\itshape\rmfamily $^*$Department of Computer Science and Engineering, The Chinese University of Hong Kong\\
\fontsize{9}{9}\selectfont\ttfamily\upshape \{$^1$yanda, $^2$jcheng, $^4$fyang, $^5$ylu, $^6$cslui, $^7$qzzhang\}@cse.cuhk.edu.hk\\
\fontsize{10}{10}\selectfont\itshape\rmfamily $^\dag$David R.\ Cheriton School of Computer Science, University of Waterloo\\
\fontsize{9}{9}\selectfont\ttfamily\upshape $^3$tozsu@uwaterloo.ca\\
\fontsize{10}{10}\selectfont\itshape\rmfamily $^+$Department of Computer Science and Engineering, The Hong Kong University of Science and Technology\\
\fontsize{9}{9}\selectfont\ttfamily\upshape $^8$wilfred@cse.ust.hk
}

\maketitle

\begin{abstract}

Pioneered by Google's Pregel, many distributed systems have been developed for large-scale graph analytics. These systems expose the user-friendly ``think like a vertex'' programming interface to users, and exhibit good horizontal scalability. However, these systems are designed for tasks where the majority of graph vertices participate in computation, but are not suitable for processing light-workload graph queries where only a small fraction of vertices need to be accessed. The programming paradigm adopted by these systems can seriously under-utilize the resources in a cluster for graph query processing. In this work, we develop a new open-source system, called {\bf Quegel}, for querying big graphs, which treats queries as first-class citizens in the design of its computing model. Users only need to specify the Pregel-like algorithm for a generic query, and Quegel processes light-workload graph queries on demand using a novel superstep-sharing execution model to effectively utilize the cluster resources. Quegel further provides a convenient interface for constructing graph indexes, which significantly improve query performance but are not supported by existing graph-parallel systems. Our experiments verified that Quegel is highly efficient in answering various types of graph queries and is up to orders of magnitude faster than existing systems.

\end{abstract}

\section{Introduction}  \label{sec:intro}

Big graphs are common in real-life applications today, for example, online social networks and mobile communication networks have billions of users, and web graphs and Semantic webs can be even bigger. Processing such big graphs typically require a special infrastructure, and the most popular ones are Pregel~\cite{pregel} and Pregel-like systems~\cite{giraph,powergraph,graphx,graphlab,gps,pregelplus}. In a Pregel-like system, a programmer \emph{thinks like a vertex} and only needs to specify the behavior of one vertex, and the system automatically schedules the execution of the specified computing logic on all vertices. The system also handles fault tolerance and scales out without extra effort from programmers.


Existing Pregel-like systems, however, are designed for \emph{heavy-weight} graph computation (i.e., analytic workloads), where the majority part of a graph or the entire graph is accessed. For example, Pregel's PageRank algorithm~\cite{pregel} accesses the whole graph in each iteration. However, many real-world applications involve various types of graph querying, whose computation is \emph{light-weight} in the sense that only a small portion of the input graph needs to be accessed. For example, in our collaboration with researchers from one of the world's largest online shopping platforms, we have seen huge demands for querying different aspects of big graphs for all sorts of analysis to boost sales and improve customer experience. In particular, they need to frequently examine the \emph{shortest-path distance} between some users in a large network extracted from their online shopping data. While Pregel's \emph{single-source shortest-path} (\emph{SSSP}) algorithm~\cite{pregel} can be applied here, much of the computation will be wasted because only those paths between the queried users are of interest. Instead, it is much more efficient to apply \emph{point-to-point shortest-path} (\emph{PPSP}) queries, which only traverse a small part of the input graph. We also worked with a large telecom operator, and our experience is that graph queries (with \emph{light-weight} workloads) are integral parts of analyzing massive mobile phone and SMS networks.

The importance of querying big graphs has also been recognized in some recent work~\cite{tutorial}, where two kinds of systems are identified: (1)~systems for offline graph analytics (such as Pregel and GraphLab) and (2)~systems for online graph querying, including Horton~\cite{SarwatEHM13pvldb}, G-SPARQL~\cite{SakrEH12pvldb} and Trinity~\cite{trinity}. However, Horton and G-SPARQL are tailor-made only for specific types of queries. Trinity supports graph query processing, but compared with Pregel, its main advantage is that it keeps the input graph in main memories so that the graph does not have to be re-loaded for each query. The Trinity paper~\cite{trinity} also argues that indexing is too expensive for big graphs and thus Trinity does not support indexing. In the VLDB 2015 conference, there is also a workshop ``Big-O(Q): Big Graphs Online Querying'', but the works presented there only study algorithms for specific types of queries. So far, there lacks a \emph{general-purpose framework} that allows users to easily design distributed algorithms for answering various types of queries on big graphs.

One may, of course, use existing vertex-centric systems to process queries on big graphs, but these systems are not suitable for processing \emph{light-weight} graph queries. To illustrate, consider processing PPSP queries on a 1.96-billion-edge Twitter graph used in our experiments. To answer one query $(s, t)$ by bidirectional breadth-first search (BiBFS) in our cluster, Giraph takes over 100 seconds, which is intolerable for a data analyst who wants to examine the distance between users in an online social network with short response time. To process queries on demand using an existing vertex-centric system, a user has the following two options: (1)~to process queries one after another, which leads to a low throughput since the communication workload of each query is usually too light to fully utilize the network bandwidth and many synchronization barriers are incurred; or (2)~to write a program to explicitly process a batch of queries in parallel, which is not easy for users and may not fully utilize the network bandwidth towards the end of the processing, since most queries may have finished their processing and only a small number of queries are still being processed. It is also not clear how to use graph indexing for query processing in existing vertex-centric systems.


To address the limitations of existing systems in querying big graphs, we developed a distributed system, called {\bf Quegel}, for large-scale graph querying. We implemented the {\em Hub$^2$-Labeling} approach~\cite{JinRYW13corr} in Quegel, and it can achieve interactive speeds for PPSP querying on the same Twitter graph mentioned above. Quegel treats queries as first-class citizens: users only need to write a Pregel-like algorithm for processing a generic query, and the system automatically schedules the processing of multiple incoming queries on demand. As a result, Quegel has a wide application scope, since any query that can be processed by a Pregel-style vertex-centric algorithm can be answered by Quegel, and much more efficiently. Under this \emph{query-centric} design, Quegel adopts a novel \emph{superstep-sharing execution model} to effectively utilize the cluster resources, and an efficient mechanism for managing vertex states that significantly reduces memory consumption. Quegel further provides a convenient interface for constructing indexes to improve query performance. To our knowledge, \emph{Quegel is the first general-purpose programming framework for querying big graphs at interactive speeds on a distributed cluster}. We have successfully applied Quegel to process five important types of graph queries (to be presented in Section~\ref{sec:app}), and Quegel achieves performance up to orders of magnitude faster than existing systems.

The rest of this paper is organized as follows. We review related work in Section~\ref{sec:related}. In Section~\ref{sec:design}, we highlight important concepts in the design of Quegel, and key implementation issues. We introduce the programming model of Quegel in Section~\ref{sec:interface}, and describe some graph querying problems as well as their Quegel algorithms in Section~\ref{sec:app}. Finally, we evaluate the performance of Quegel in Section~\ref{sec:results} and conclude the paper in Section~\ref{sec:conclude}.

\section{Related Work}\label{sec:related}
We first review existing vertex-centric graph-parallel systems. We consider an input graph $G=(V, E)$ stored on \emph{Hadoop distributed file system} (\emph{HDFS}), where each vertex $v\in V$ is associated with its adjacency list (i.e., $v$'s neighbors). If $G$ is undirected, we denote $v$'s neighbors by $\Gamma(v)$, while if $G$ is directed, we denote $v$'s in-neighbors and out-neighbors by $\Gamma_{in}(v)$ and $\Gamma_{out}(v)$, respectively. Each vertex $v$ also has a value $a(v)$ storing $v$'s vertex value. Graph computation is run on a cluster of workers, where each worker is a computing thread/process, and a machine may run multiple workers.

\vspace{2mm}

\noindent{\bf Pregel~\cite{pregel}.} Pregel adopts the \emph{bulk synchronous parallel} (\emph{BSP}) model. It distributes vertices to workers in a cluster, where each vertex is associated with its adjacency list. A Pregel program computes in iterations, where each iteration is called a superstep. Pregel requires users to specify a \emph{user-defined function} (\emph{UDF}) {\em compute}(.). In each superstep, each active vertex $v$ calls {\em compute}({\em msgs}), where {\em msgs} is the set of incoming messages sent from other vertices in the previous superstep. In $v$.{\em compute}({\em msgs}), $v$ may process {\em msgs} and update $a(v)$, send new messages to other vertices, and vote to halt (i.e., deactivate itself). A halted vertex is reactivated if it receives a message in a subsequent superstep. The program terminates when all vertices are deactivated and no new message is generated. Finally, the results (e.g., $a(v)$) are dumped to HDFS.

Pregel also allows users to implement an aggregator for global communication. Each vertex can provide a value to an aggregator in {\em compute}(.) in a superstep. The system aggregates those values and makes the aggregated result available to all vertices in the next superstep.

\vspace{2mm}


\noindent{\bf Distributed Vertex-Centric Systems.} Many Pregel-like systems have been developed, including Giraph~\cite{giraph}, GPS~\cite{gps}, GraphX~\cite{graphx}, and Pregel+~\cite{pregelplus}. New features are introduced by these systems, for example, GPS proposed to mirror high-degree vertices on other machines, and Pregel+ proposed the integration mirroring and message combining as well as a request-respond mechanism, to reduce communication workload. While these systems strictly follow the synchronous data-pushing model of Pregel, GraphLab~\cite{graphlab} adopts an asynchronous data-pulling model, where each vertex actively pulls data from its neighbors rather than passively receives messages. A subsequent version of GraphLab, called PowerGraph~\cite{powergraph}, partitions the graph by edges rather than by vertices to achieve more balanced workload. While the asynchronous model leads to faster convergence for some tasks like random walk, \cite{ourExp} and~\cite{tamerExp} reported that GraphLab's asynchronous mode is generally slower than synchronous execution mainly due to the expensive cost of locking/unlocking.


\vspace{2mm}


\noindent{\bf Single-PC Vertex-Centric Systems.} There are also other vertex-centric systems, such as GraphChi~\cite{graphchi} and X-Stream~\cite{xstream}, designed to run on a single PC by manipulating a big graph on disk. However, these systems need to scan the whole graph on disk once for each iteration of computation even if only a small fraction of vertices need to perform computation, which is inefficient for light-weight querying workloads.


\vspace{2mm}

\noindent{\bf Weaknesses of Existing Systems for Graph Querying.} In our experience of working with researchers in e-commerce companies and telecom operators, we found that existing vertex-centric systems cannot support query processing efficiently nor do they provide a user-friendly programming interface to do so. If we write a vertex-centric algorithm for a generic query, we have to run a job for every incoming query. As a result, each superstep transmits only the few messages of one light-weight query which cannot fully utilize the network bandwidth. Moreover, there are a lot of synchronization barriers, one for each superstep of each query, which is costly. Moreover, some systems such as Giraph bind graph loading with graph computation (i.e., processing a query in our context) for each job, and the loading time can significantly degrade the performance.

An alternative to the one-query-at-a-time approach is to hard code a vertex-centric algorithm to process a batch of $k$ queries, where $k$ can be an input argument. However, in the {\em compute}(.) function, one has to differentiate the incoming messages and/or aggregators of different queries and update $k$ vertex values accordingly. In addition, existing vertex-centric framework checks the stop condition for the whole job, and users need to take care of additional details such as when a vertex can be deactivated (e.g., when it should be halted for all the $k$ queries), which should originally be handled by the system itself. More critically, the one-batch-at-a-time approach does not solve the problem of low utilization of network bandwidth, since in later stage when most queries finish their processing, only a small number of queries (or stragglers) are still being processed and hence the number of messages generated is too small to sufficiently utilize the network bandwidth.


The single-PC systems are clearly not suitable for light-weight querying workloads since they need to scan the whole graph on disk once for each iteration. Other existing graph databases such as Neo4j~\cite{neo4j} and HyperGraphDB~\cite{Iordanov10waim} support basic graph operations and simple graph queries, but they are not designed to handle big graphs. Our experiments also verified the inefficiency of single-PC systems and graph databases in querying big graphs (see Section~\ref{sec:results}). There are other systems, e.g., the block-centric system Blogel~\cite{blogel} and a recent general-purpose system Husky~\cite{YangLC16pvldb}, which achieve remarkable performance on offline graph analytics, but are not designed for graph querying.

The above discussion motivates the need of a general-purpose graph processing system that treats queries as first citizens, which provides a user-friendly interface so that users can write their program easily for one generic query and the system processes queries on demand efficiently. Our Quegel system, to be presented in the following sections, fulfils this need.

\section{The Quegel System}\label{sec:design}
A Quegel program starts by loading the input graph $G$, i.e., distributing vertices into the main memory of different workers in a cluster. If users enable indexing, a local index will be built from the vertices of each worker. After $G$ is loaded (and index is constructed), Quegel receives and processes incoming queries using the computing logic specified by a vertex UDF {\em compute}(.) as in Pregel. Users may type their queries from a client console, or submit a batch of queries with a file. After a query is evaluated, users may specify Quegel to print the answer to the console, or to dump the answer to HDFS if its size is large (e.g., the answer contains many subgraphs).

\subsection{Execution Model: Superstep-Sharing}

To address the weaknesses of existing systems presented in Section~\ref{sec:related}, we need to consider a new computation model. We first present the hardness of querying a big graph in general, which influences the design of our model.

\vspace{2mm}

\noindent{\bf Hardness of Big Graph Querying and Our Design Objective.} We consider the processing of a large graph that is stored in distributed sites, so that the processing of each query requires network communication. Since the message transmission of each superstep incurs round-trip delay, it is difficult (if not unrealistic) for distributed vertex-centric computation (e.g., on $k$ machines) to achieve response time comparable to that of single-machine algorithms on a smaller graph (e.g., $k$ times smaller). Therefore, our goal is to answer a query in interactive speed, e.g., in a second to at most a few seconds depending on the complexity of processing a given query. We remark that even in CANDS~\cite{cands}, a specialized distributed system dedicated for shortest path querying on big graphs, a query can take many seconds to answer, while as we shall see in Section~\ref{sec:results}, our general-purpose Quegel system can process multiple PPSP queries per second on a graph with billions of edges.

Moreover, due to the sheer size of a big graph, the total workload of a batch of queries can be huge even if each query accesses just a fraction of the graph. We remark that the workload of distributed graph computation is significantly different from traditional database applications. For example, to query the balance of a bank account, the balance value can be quickly accessed from a centralized account table using a B$^+$-tree index based on the account number, and it is possible to achieve both high throughput and low latency. However, in distributed graph computation, the complicated topology of connections among vertices (which are not present among bank accounts) results in higher-complexity algorithms and heavier workloads. Specifically, due to the poor locality of graph data, each query usually accesses vertices spreading through the whole big graph in distributed sites, and vertices need to communicate with each other through the network.

The above discussion shows that there is a latency-throughput tradeoff where one can only expect either interactive speed or high throughput but not both. As a result, our \emph{design objective} focuses on developing a model for the following two scenarios of querying big graphs, both of which are common in real life applications.

\vspace{2mm}

\noindent{\bf Scenario~(i): Interactive Querying}, where a user interacts with Quegel by submitting a query, checking the query results, refining the query based on the results and re-submitting the refined query, until the desired results are obtained. As an example, a data analyst may use interactive PPSP queries to examine the distance between two users of interest in a social network. Another example is given by the XML keyword querying application (to be presented in Section~\ref{ssec:xml}). In such applications, there are only one or several users (e.g., a data scientist) analyzing a big graph by posing interactive queries, but each query should be answered in a second or several seconds. No existing vertex-centric system can achieve such query latency on a big graph.

\vspace{2mm}

\noindent{\bf Scenario~(ii): Batch Querying}, where batches of queries are submitted to Quegel, and they need to be answered within a reasonable amount of time. An example of batch querying is given by the vertex-pair sampling application mentioned in Section~\ref{sec:intro} for estimating graph metrics, where a large number of PPSP queries need to be answered. Quegel achieves throughput 186 and 38.6 times higher than Giraph and GraphLab for processing PPSP queries, and thus allows the graph metrics to be estimated more accurately.

\vspace{2mm}

\noindent{\bf Superstep-Sharing Model.} We propose a \emph{superstep-sharing execution model} to meet the requirements of both interactive querying and batch querying. Specifically, Quegel processes graph queries in iterations called  \textbf{super-rounds}. In a super-round, every query that is currently being processed proceeds its computation by one superstep; while from the perspective of an individual query, Quegel processes it superstep by superstep as in Pregel. Intuitively, a super-round in Quegel is like \emph{many queries sharing the same superstep}. For a query $q$ whose computation takes $n_q$ supersteps, Quegel processes it in $(n_q+1)$ super-rounds, where the last super-round reports or dumps the results of $q$.


Quegel allows users to specify a capacity parameter $C$, so that in any super-round, there are at most $C$ queries being processed. New incoming queries are appended to a query queue, and at the beginning of a super-round, Quegel fetches as many queries from the queue as possible to start their processing, as long as the capacity constraint $C$ permits. During the computation of a super-round, different workers run in parallel, while each worker processes (its part of) the evaluation of the queries serially. And for each query $q$, if $q$ has not been evaluated, a worker serially calls {\em compute}(.) on each of its vertices that are activated by $q$; while if $q$ has already finished its evaluation, the worker reports or dumps the query results, and releases the resources consumed by $q$.

For the processing of each query, the supersteps are numbered. Different queries may have different superstep number in the same super-round, for example, if the queries enter the system in different super-rounds. Messages (and aggregators) of all queries are synchronized together at the end of a super-round, to be used by the next super-round.

For interactive querying where queries are posed and processed in sequence, the superstep-sharing model processes each individual query with all the cluster resources just as in Pregel. However, since Quegel decouples the costly graph loading and dumping from query processing, and supports convenient construction and adoption of graph indexes, the query latency is significantly reduced.

\begin{figure}[!t]
    \centering
    \includegraphics[width=\columnwidth]{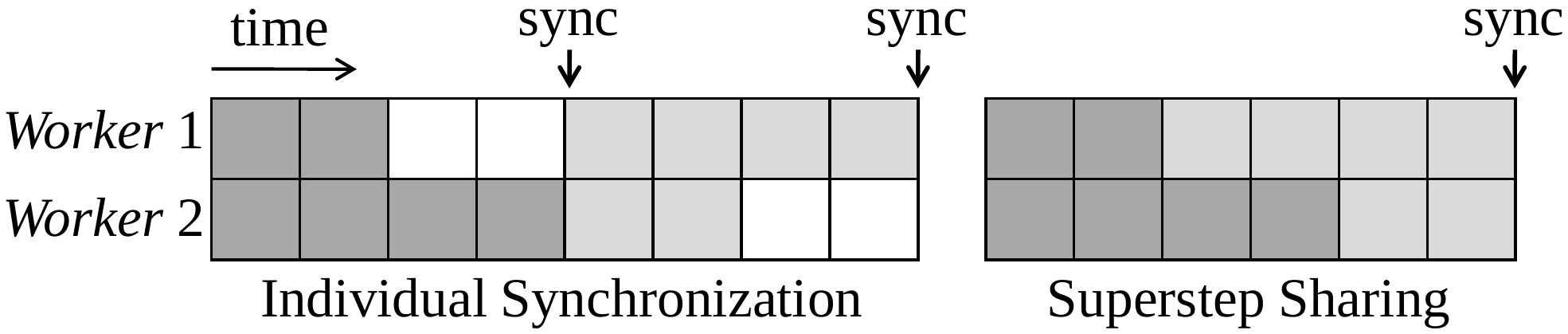}
    \caption{Load balancing}\label{balance}
\end{figure}

\begin{figure*}[!t]
    \centering
    \includegraphics[width=2\columnwidth]{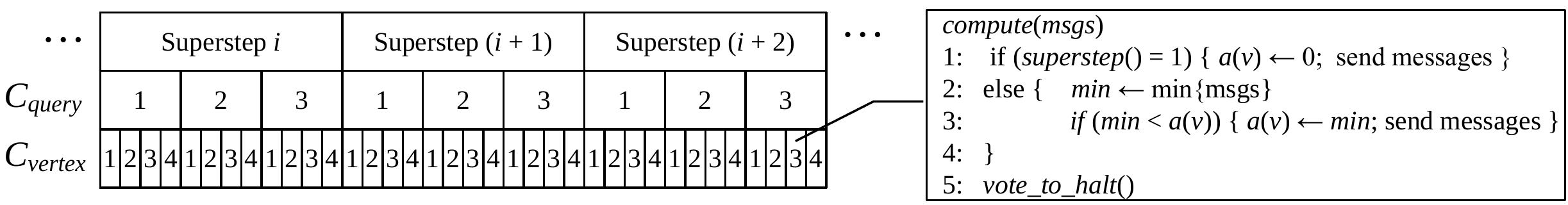}
    \caption{Illustration of context objects}\label{context}
\end{figure*}

For batch querying, while the workload of each individual query is light, superstep-sharing combines the workloads of up to $C$ queries as one batch in each super-round to achieve higher resource utilization. Compared with answering each query independently as in existing graph-parallel systems, Quegel's superstep-sharing model supports much more efficient query processing since only one message (and/or aggregator) synchronization barrier is required in each super-round instead of up to $C$ synchronization barriers. We remark that the synchronization cost is relatively significant compared with the light workload of processing each single query. In addition, by sending the messages of many queries in one batch, superstep-sharing also better utilizes the network bandwidth.

Superstep-sharing also leads to more balanced workload. As an illustration, Figure~\ref{balance} shows the execution of two queries for one superstep in a cluster of two workers. The first query (darker shading) takes 2 time units on Worker~1 and 4 time units on Worker~2, while the second query (lighter shading) takes 4 time units on Worker~1 and 2 time units on Worker~2. When the queries are processed individually, the first query needs to be synchronized before the second query starts to be processed. Thus, 8 time units are required in total. Using superstep-sharing, only one synchronization is needed at the end of the super-round, thus requiring only 6 time units.

One issue that remains is how to set the capacity parameter $C$. Obviously, the larger the number of queries being simultaneously processed, the more fully is the network bandwidth utilized. But the value of $C$ should be limited by the available RAM space. The input graph consumes $O(|V|+|E|)$ RAM space, while each query $q$ consumes $O(|V_q|)$ space, where $V_q$ denotes the set of vertices accessed by $q$. Thus, $O(|V|+|E|+C|V_q|)$ should not exceed the available RAM space, though in most case this is not a concern as $|V_q|\ll|V|$. While setting $C$ larger tends to improve the throughput, the throughput converges when the network bandwidth is saturated. In a cluster such as ours which is connected by Gigabit Ethernet, we found that the throughput usually converges when $C$ is increased to 8 (for the graph queries we tested), which indicates that Quegel has already fully utilized the network bandwidth and shows the high complexity of querying a big graph.

\subsection{System Design}\label{ssec:design}

Quegel manages three kinds of data: {\bf(i)~V-data}, whose value only depends on a vertex $v$, such as $v$'s adjacency list. {\bf(ii)~VQ-data}, whose value depends on both a vertex $v$ and a query $q$. For example, the vertex value $a(v)$ is query-dependent: in a PPSP query $q=(s, t)$, $a(v)$ keeps the estimated value of the shortest distance from $s$ to $v$, denoted by $d(s, v)$, whose value depends on the source vertex $s$. As $a(v)$ is  w.r.t.\ a query $q$, we use $a_q(v)$ to denote ``$a(v)$ w.r.t.\ $q$''. Other examples of VQ-data include the active/halted state of a vertex $v$, and the incoming message buffer of $v$ (i.e., input to $v.${\em compute}(.)). {\bf(iii)~Q-data}, whose value only depends on a query $q$. For example, at any moment, each query $q$ has a unique superstep number. Other examples of Q-data include the query content (e.g., $(s, t)$ for a PPSP query), the outgoing message buffers, aggregated values, and control information that decides whether the computation should terminate.

Let $Q=\{q_1, \ldots, q_k\}$ be the set of queries currently being processed by Quegel, and let $id(q_i)$ be the query ID of each $q_i\in Q$.

In Quegel, each worker maintains a hash table $HT_Q$ to keep the Q-data of each query in $Q$. The Q-data of a query $q_i$ can be obtained from $HT_Q$ by providing the query ID $id(q_i)$, and we denote it by $HT_Q[q_i]$. When a new query $q$ is fetched from the query queue to start its processing at the beginning of a super-round, the Q-data of $q$ is inserted into $HT_Q$ of every worker; while after $q$ reports or dumps its results at superstep $(n_q+1)$, the Q-data of $q$ is removed from $HT_Q$ of every worker.

Each worker $W$ also maintains an array of vertices, {\em varray}, each element of which maintains the V-data and VQ-data of a vertex $v$ that is distributed to $W$. The VQ-data of a vertex $v$ is organized by a look-up table $LUT_v$, where the VQ-data related to a query $q_i$ can be obtained by providing the query ID $id(q_i)$, and we denote it by $LUT_v[q_i]$. Since every vertex $v$ needs to maintain a table $LUT_v$, we implement it using a space-efficient balanced binary search tree rather than a hash table. The data kept by each table entry $LUT_v[q]$ include the vertex value $a_q(v)$, the active/halted state of $v$ (in $q$), and the incoming message buffer of $v$ (for $q$).

Unlike the one-batch-at-a-time approach of applying existing vertex-centric systems, where each vertex $v$ needs to maintain $k$ vertex values no matter whether it is accessed by a query, we design Quegel to be more space efficient. We require that {\it a vertex $v$ is allocated a state for a query $q$ only if $q$ accesses $v$ during its processing}, which is achieved by the following design. When vertex $v$ is activated for the first time during the processing of $q$, the VQ-data of $q$ is initialized and inserted into $LUT_v$. After a query $q$ reports or dumps its results at superstep $(n_q+1)$, the VQ-data of $q$ (i.e., $LUT_v[q]$) is removed from $LUT_v$ of every vertex $v$ in $G$.

Each worker also maintains a hash table $HT_V$, such that the position of a vertex element $v$ in {\em varray} can be obtained by providing the vertex ID of $v$. We denote the obtained vertex element by $HT_V[v]$. The table $HT_V$ is useful in two places: (1)~when a message targeted at vertex $v$ is received, the system will obtain the incoming message buffer of $v$ from {\em varray}$[pos]$ where $pos$ is computed as $HT_V[v]$, and then append the message to the buffer; (2)~when an initial vertex $v$ is activated using its vertex ID at the beginning of a query, the system will initialize the VQ-data of $v$ for $q$, and insert it into $LUT_v$ which is obtained from {\em varray}$[pos]$ where $pos$ is computed as $HT_V[v]$. We shall see how users can activate the (usually small) initial set of vertices in Quegel for processing without scanning all vertices in Section~\ref{sec:interface}.

An important feature of Quegel is that, it only requires a user to specify the computing logic for a generic vertex and a generic query; the processing of concrete queries is handled by Quegel and is totally transparent to users. For this purpose, each worker $W$ maintains two \emph{global context objects}: (i)~query context $C_{query}$, which keeps the Q-data of the query that $W$ is processing; and (ii)~vertex context $C_{vertex}$, which keeps the VQ-data of the current vertex that $W$ is processing for the current query. In a super-round, when a worker starts to process each query $q_i$, it first obtains $HT_Q[q_i]$ and assigns it to $C_{query}$, so that when a user accesses the Q-data of the current query in UDF {\em compute}(.) (e.g., to get the superstep number or to append messages to outgoing message buffers), the system will access $C_{query}$ directly without looking up from $HT_Q$. Moreover, during the processing of $q_i$, and before the worker calls {\em compute}(.) on each vertex $v$, it first obtains $LUT_v[q_i]$ and assigns it to $C_{vertex}$, so that any access or update to the VQ-data of $v$ in {\em compute}(.) (e.g., obtaining $a_q(v)$ or letting $v$ vote to halt) directly operates on $C_{vertex}$ without looking up from $LUT_v$.

As an illustration, consider the example shown in Figure~\ref{context}, where there are 3 queries being evaluated and the computation proceeds for 3 supersteps. Moreover, we assume that 4 vertices call {\em compute}(.) in each superstep of each query. As an example, when processing a superstep~$(i+2)$, $C_{query}$ is set to $HT_Q[q_3]$ before evaluating $v_1$ for $q_3$; and when the evaluation arrives at $v_3$, $C_{vertex}$ is set to $LUT_{v_3}[q_3]$ before $v_3.${\em compute}(.) is called. Figure~\ref{context} also shows a simplified code of {\em compute}(.) for shortest path computation, and inside $v_3.${\em compute}(.) for $q_3$, $a(v)$ is accessed once in Line~1 and twice in Line~3, all of which use the value $a_{q_3}(v)$ stored in $C_{vertex}=LUT_{v_3}[q_3]$ directly; while Line~1 accesses the superstep number which is obtained from $C_{query}=HT_Q[q_3]$ directly.

One benefit of using the context objects $C_{vertex}$ and $C_{query}$ is that, due to the access pattern locality of superstep-sharing, {\it repetitive lookups of tables $HT_Q$ and $LUT_{v}$ are avoided}. Another benefit is that, {\it users can write their program exactly like in Pregel} (e.g., to access $a(v)$ and superstep number) and the processing of concrete queries is transparent to users.

\section{Programming Interface}\label{sec:interface}
The programming interface of Quegel incorporates many unique features designed for querying workload. For example, the interface allows users to construct distributed graph indexes at graph loading. The interface also allows users to activate only an initial (usually small) set of vertices, denoted by $V^I_q$, for processing a query $q$ without checking all vertices. Note that we cannot activate $V^I_q$ during graph loading because $V^I_q$ depends on each incoming query $q$.

Quegel defines a set of base classes, each of which is associated with some template arguments. To write an application program, a user only needs to (1)~subclass the base classes with the template arguments properly specified, and to (2)~implement the UDFs according to the application logic. We now describe these base classes.

\vspace{2mm}

\noindent{\bf Vertex Class.} As in Pregel, the {\em Vertex} class has a UDF {\em compute}(.) for users to specify the computing logic. In {\em compute}(.), a user may call {\em get\_query}() to obtain the content of the current query $q_{cur}$. A user may also access other Q-data in {\em compute}(.), such as getting $q_{cur}$'s superstep number, sending messages (which appends messages to $q_{cur}$'s outgoing message buffers), and getting $q_{cur}$'s aggregated value from the previous superstep. Quegel also allows a vertex to call {\em force\_terminate}() to terminate the computation of $q_{cur}$ at the end of the current superstep. All these operations access the Q-data fields from $C_{query}$ directly.

The vertex class of Quegel is defined as {\em Vertex}$<$$I, V^Q, V^V, M, Q$$>$, which has five template arguments: (1)~$<$$I$$>$ specifies the type (e.g., {\tt int}) of the ID of a vertex (which is V-data). (2)~$<$$V^Q$$>$ specifies the type of the query-dependent attribute of a vertex $v$, i.e., $a_q(v)$ (which is VQ-data). (3)~$<$$V^V$$>$ specifies the type of the query-independent attribute of a vertex $v$, denoted by $a^V(v)$ (which is V-data). We do not hard-code the adjacency list structure in order to provide more flexibility. For example, a user may define $a^V(v)$ to include two adjacency lists, one for in-neighbors and the other for out-neighbors, which is useful for algorithms such as bidirectional BFS. Other V-data can also be included in $a^V(v)$, such as vertex labels used for search space pruning in some query processing algorithms. (4)~$<$$M$$>$ specifies the type of the messages that are exchanged between vertices. (5)~$<$$Q$$>$ specifies the type of the content of a query. For example, for a PPSP query, $<$$Q$$>$ is a pair of vertex IDs indicating the source and target vertices. In {\em compute}(.), a user may access $a^V(v)$ by calling {\em value}$()$, and access $a_q(v)$ by calling {\em qvalue}$()$.

Suppose that a set of $k$ queries, $Q$, is being processed, then each vertex conceptually has $k$ query-dependent attributes $a_q(v)$, one for each query $q\in Q$. Since a query normally only accesses a small fraction of all the vertices, to be space-efficient, Quegel allocates space to $a_q(v)$ as well as other VQ-data only at the time when the vertex is first accessed during the processing of $q$. Accordingly, Quegel provides a UDF {\em init\_value}($q$) for users to specify how to initialize $a_q(v)$ when $v$ is first accessed by $q$. For example, for a PPSP query $q=(s, t)$, where $a_q(v)$ keeps the estimated value of $d(s, v)$, one may implement {\em init\_value}($s, t$) as follows: if $v = s$, $a_q(v) \gets 0$; else, $a_q(v) \gets \infty$. The state of $v$ is always initialized to be active by the system, since when the space of the state is allocated, $v$ is activated for the first time and should participate in the processing of $q$ in the current superstep. Function {\em init\_value}($q$) is the only UDF of the {\em Vertex} class in addition to {\em compute}(.).

\vspace{2mm}

\noindent{\bf Worker Class.} The {\em Vertex} class presented above is mainly for users to specify the graph computation logic. Quegel provides another base class, {\em Worker}$<$$T_{vtx}, T_{idx}$$>$, for specifying the input/output format and for executing the computation of each worker. The template argument $<$$T_{vtx}$$>$ specifies the user-defined subclass of {\em Vertex}. The template argument $<$$T_{idx}$$>$ is optional, and if distributed indexing (to be introduced shortly) is enabled, $<$$T_{idx}$$>$ specifies the user-defined index class.

The {\em Worker} class has a function {\em run}({\em param}), which implements the execution procedure of Quegel as described at the beginning of Section~\ref{sec:design}. After users define their subclasses to implement the computing logic, they call {\em run}({\em param}) to start a Quegel job. Here, {\em param} specifies job parameters such as the HDFS path of the input graph $G$. During the execution, we allow each query to change $a^V(v)$ of a vertex $v$, and when a user closes the Quegel program from the console, he/she may specify Quegel to save the changed graph (V-data only) to HDFS, before freeing the memory space consumed by $G$.

The {\em Worker} class has four formatting UDFs, which are used (1)~to specify how to parse a line of the input file into a vertex of $G$ in main memory, (2)~to specify how to parse a query string (input by a user from the console or a file) into the query content of type $<$$Q$$>$, (3)~to specify how to write the information of a vertex $v$ (e.g., $a_q(v)$) to HDFS after a query is answered, and (4)~to specify how to write the changed V-data of a vertex $v$ to HDFS when a Quegel job terminates. The last UDF is optional, and is only useful if users enable the end-of-job graph dumping.

Quegel allows each worker to construct a local index from its loaded vertices before query processing begins. We illustrate this process by considering a vertex-labeled graph $G$ where each vertex $v$ contains text $\psi(v)$, and show how to construct an inverted index on each worker $W$, so that given a keyword $k$, it returns a list of vertices on $W$ whose text contains $k$. This kind of index is useful in XML keyword search~\cite{LiuC08pvldb,ZhouBWLCLG12icde}, subgraph pattern matching~\cite{ChengYDYW08icde,GaoZZY14aicde}, and graph keyword search~\cite{HeWYY07sigmod,QinYCCZL14sigmod}. Specifically, recall that each worker in Quegel maintains its vertices in an array {\em varray}. If indexing is enabled, a UDF {\em load2Idx}($v, pos$) will be called to process each vertex $v$ in {\em varray} immediately after graph loading, where $pos$ is $v$'s position in {\em varray}. To construct inverted indexes in Quegel, a user may specify $<$$T_{idx}$$>$ as a user-defined inverted index class, and implement {\em load2Idx}($v, pos$) to add $pos$ to the inverted list of each keyword $k$ in $\psi(v)$. There are also indices that cannot be constructed simply from local vertices, and we shall see how to handle such an application in Quegel in Section~\ref{ssec:ppsp}.

When a query is first scheduled for processing, each worker calls a UDF {\em init\_activate}() to activate only the relevant vertices specified by users. For example, in a PPSP query $(s, t)$, only $s$ and $t$ are activated initially; while for querying a vertex-labeled graph, only those vertices whose text contain at least one keyword in the query are activated. Inside {\em init\_activate}(), one may call {\em get\_vpos}({\em vertexID}) to get the position $pos$ of a vertex in {\em varray} (which actually looks up the hash table $HT_V$ of each worker), and then call {\em activate}($pos$) to activate the vertex. For example, to activate $s$ in a PPSP query $(s, t)$, a user may specify {\em init\_activate}() to first call {\em get\_vpos}($s$) to return $s$'s position $pos_s$. If $s$ is on the current worker, $pos_s$ will be returned and one may then call {\em activate}($pos_s$) to activate $s$ in {\em init\_activate}(). If $s$ is not on the current worker, {\em get\_vpos}($s$) returns -1 and no action needs to be performed in {\em init\_activate}(). For querying a vertex-labeled graph, a user may specify {\em init\_activate}() to first get the positions of the keyword-matched vertices from the inverted index, and then activate them using {\em activate}($pos$).

\vspace{2mm}

\noindent{\bf Other Base Classes.} Quegel also provides other base classes such as {\em Combiner} and {\em Aggregator}, for which users can subclass them to specify the logic of message combiner~\cite{pregel} and aggregator~\cite{pregel}.

\section{Applications}\label{sec:app}
To demonstrate the generality of Quegel's computing model for querying big graphs, we have implemented distributed
algorithms for five important types of graph queries in Quegel, including (1)~PPSP queries, (2)~XML keyword queries, (3)~terrain shortest path queries, (4)~point-to-point (P2P) reachability queries, and (5)~graph keyword queries. Among them, (1), (3) and~(4) only care about the graph topology, while (2) and~(5) also care about the text information on vertices and edges. We now present the five applications and their Quegel solutions.

\subsection{PPSP Queries}\label{ssec:ppsp}
We consider a PPSP query defined as follows. Given two vertices $s$ and $t$ in an unweighted graph $G=(V, E)$, find the minimum number of hops from $s$ to $t$ in $G$, denoted by $d(s, t)$. We focus on unweighted graphs since most large real graphs (e.g., social networks and web graphs) are unweighted. Moreover, we are only interested in reporting $d(s, t)$, although our algorithms can be easily modified to output the actual shortest path(s).

\subsubsection{Algorithms without Indexing}\label{sssec:no_idx}
\noindent{\bf Breadth-First Search (BFS).} The simplest way of answering a PPSP query $q=(s, t)$ is to perform BFS from $s$, until the search reaches $t$. In this algorithm, $a_q(v)$ is specified to be the current estimation of $d(s, v)$, and we use $d(s, v)$ to denote $a_q(v)$ in our discussion for simplicity. The UDF {\em init\_activate}() of user-defined {\em Worker} subclass should activate $s$ at the beginning of processing $q$. The vertex UDF $v.${\em init\_value}$(s, t)$ should set $d(s, v)$ to $0$ if $v=s$, and to $\infty$ otherwise. Note that $v$ calls {\em init\_value}(.) when $v$ is first activated during the processing of $q$, either by {\em init\_activate}() or because some vertex sends $v$ a message.

The vertex UDF $v$.{\em compute}$(.)$ is implemented as follows. Let $step_q$ be the superstep number of $q$. If $step_q=1$, then $v$ must be $s$ since only $s$ is activated by {\em init\_activate}(); $s$ broadcasts messages to its out-neighbors to activate them, and then votes to halt. If $step_q>1$, one of the following is performed: (i)~if $d(s, v)=\infty$, then $v$ is visited by the BFS for the first time; in this case, $v$ sets $d(s, v)\gets step_q-1$, broadcasts messages to activate $v$'s out-neighbors and votes to halt; if $v=t$, $v$ also calls {\em force\_terminate}() to terminate query processing as $d(s, t)$ has been computed; (ii)~if $d(s, v)\neq\infty$, then $v$ has been activated by $q$ before, and hence $v$ votes to halt directly. Finally, only $t$ reports $a_q(t)=d(s, t)$ on the console and nothing is dumped to HDFS.

\vspace{2mm}

\noindent{\bf Bidirectional BFS (BiBFS).} A more efficient algorithm is to perform forward BFS from $s$ and backward BFS from $t$ until a vertex $v$ is visited in both directions, and we say that $v$ is {\em bi-reached} in this case. Let $C$ be the set of bi-reached vertices when BiBFS stops, then $d(s, t)$ is given by $\min_{v\in C}\{d(s, v)+d(v, t)\}$. We take the minimum since when BiBFS stops at iteration~$i$, $(d(s, v)+d(v, t))$ for a vertex $v\in C$ may be either $(2i-1)$ or $2i$.

The Quegel algorithm for BiBFS is similar to that for BFS, with the following changes. The query-dependent vertex attribute $a_q(v)$ now keeps a pair $(d(s, v), d(v, t))$. The vertex UDF $v.${\em init\_value}$(s, t)$ sets $d(s, v)$ to $0$ if $v=s$, and to $\infty$ otherwise; and it sets $d(v, t)$ to $0$ if $v=t$, and to $\infty$ otherwise. Both $s$ and $t$ are activated by {\em init\_activate}() initially, and two types of messages are used in order to perform forward BFS and backward BFS in parallel without interfering with each other. In $v$.{\em compute}$(.)$, if both $d(s, v)\neq\infty$ and $d(v, t)\neq\infty$, $v$ should call {\em force\_terminate}() since $v$ is bi-reached. Then, an aggregator is used to collect the distance $(d(s, v)+d(v, t))$ of each $v\in C$, and to obtain the smallest one as $d(s, t)$ for reporting.

BiBFS may be inferior to BFS in the following situation. Suppose that $G$ is undirected, and $s$ is in a small connected component (CC) while $t$ is in another giant CC. BFS will terminate quickly after all vertices in the small CC are visited, while BiBFS continues computation until all vertices in the giant CC are also visited. To solve this problem, we use aggregator to compute the numbers of messages sent by the forward BFS and the backward BFS in each superstep, respectively. If the number of messages sent in either direction is 0, the aggregator calls {\em force\_terminate}() and reports $d(s, t)=\infty$.

\subsubsection{Hub$^2$: An Algorithm with Indexing}
Many big graphs exhibit skewed degree distribution, where some vertices (e.g., celebrities in a social network) connect to a large number of other vertices. We call such vertices as \emph{hubs}. During BFS, visiting a hub results in visiting a large number of vertices at the next step, rendering BFS or BiBFS inefficient. {\em Hub$^2$-Labeling} (abbr.\ {\em Hub$^2$})~\cite{JinRYW13corr} was proposed to address this problem. We present a distributed implementation of {\em Hub$^2$} in Quegel for answering PPSP queries. We first consider undirected graphs and then extend the method to directed graphs.

{\em Hub$^2$} picks $k$ vertices with the highest degrees as the hubs. Let us denote the set of hubs by $H$, {\em Hub$^2$} pre-computes the pairwise distance between any pair of hubs in $H$. {\em Hub$^2$} also associates each vertex $v\notin H$ with a list of hubs, $H_v\subseteq H$, called \emph{core-hubs}, and pre-computes $d(v, h)$ for each core-hub $h\in H_v$. Here, a hub $h\in H$ is a core-hub of $v$, iff no other hub exists on any shortest path between $v$ and $h$. Formally, each vertex $v\in V$ maintains a list $L(v)$ of hub-distance labels defined as follows: (i)~if $v\in H$, $L(v)=\{\langle u, d(v, u)\rangle\ |\ u\in H\}$; (ii)~if $v\in (V-H)$, $L(v)=\{\langle u, d(v, u)\rangle\ |\ u\in H_v\}$.

Given a PPSP query $q=(s, t)$, an upperbound of $d(s, t)$ can be derived from the vertex labels. For ease of presentation, we only present the algorithm for the case where neither $s$ nor $t$ is a hub, while algorithms for the other cases can be similarly derived. Specifically, $d(s, t)$ is upperbounded by $d_{ub}=\min_{h_s\in H_s, h_t\in H_t}\{d(s, h_s)+d(h_s, h_t)+d(h_t, t)\}$. Obviously, if there exists a shortest path $P$ from $s$ to $t$ that passes at least one hub (note that we allow $h_s=h_t$), then $d_{ub}$ is exactly $d(s, t)$. However, the shortest path $P'$ from $s$ to $t$ may not contain any hub, and thus we still need to perform BiBFS from $s$ and $t$. Note that any edge $(u, v)$ on $P'$ satisfies $u, v\not\in H$, and thus {\em we need not continue BFS from any hub}. In other words, BiBFS is performed on the subgraph of $G$ induced by $(V-H)$, which does not include high-degree hubs.

\vspace{2mm}

\noindent{\bf Algorithm for Querying.} We now present the UDF {\em compute}(.), which applies {\em Hub$^2$} to process PPSP queries. We first assume that $L(v)$ for each vertex $v$ is already computed (we will see how to compute $L(v)$ shortly), and that $v$ keeps the query-independent attribute $a^V(v)=(\Gamma(v), L(v))$. The algorithm for BiBFS is similar to the one discussed before, with the following changes: (i)~whenever forward or backward BFS visits a hub $h$, $h$ votes to halt directly; and (ii)~once a vertex $v \notin H$ is bi-reached, $v$ calls {\em force\_terminate}() to terminate the computation, and reports $\min_{v\in (C-H)}\{d(s, v)+d(v, t)\}$. Moreover, the BiBFS should terminate earlier if the superstep number reaches $i=(1+\lfloor\frac{d_{ub}}{2}\rfloor)$ (even if no vertex is bi-reached), and $d(s, t)=d_{ub}$ is reported. This is because, a non-hub vertex $v$ that is bi-reached at superstep~$i$ or later would report $d(s, v)+d(v, t)\geq(2i-1)$, which cannot be smaller than $d_{ub}$.

We obtain $d_{ub}$ in the first two supersteps: in superstep~1, only $s$ and $t$ have been activated by {\em init\_activate}(); $s$ sends each core-hub $h_s\in H_s$ a message $\langle d(s, h_s)\rangle$ (obtained from $L(s)$), while $t$ provides $L(t)$ to the aggregator. In superstep~2, each vertex $h_s\in H_s$ receives message $d(s, h_s)$ from $s$, and obtains $L(t)$ from the aggregator. Then, $h_s$ evaluates $\min_{h_t\in H_t}\{d(s, h_s)+d(h_s, h_t)+d(h_t, t)\}$, where $d(h_s, h_t)$ is obtained from $L(h_s)$ and $d(h_t, t)$ is obtained from $L(t)$, and provides the result to the aggregator. The aggregator takes the minimum of the values provided by all $h_s\in H_s$, which gives $d_{ub}$.

\vspace{2mm}

\noindent{\bf Algorithm for Indexing.} The above algorithm requires that each vertex $v$ stores $L(v)$ in $a^V(v)$. We now consider how to pre-compute $L(v)$ in Quegel. This indexing procedure can be accomplished by performing $|H|$ BFS operations, each starting from a hub $h\in H$. Interestingly, if we regard each BFS operation from a hub $h$ as a BFS query $\langle h\rangle$ in Quegel, then the entire procedure can be formulated as an independent Quegel job with the query set $\{\langle h\rangle\ |\ h\in H\}$.

We process a BFS query $\langle h\rangle$ in Quegel as follows. The query-dependent attribute of a vertex $v$ is defined as $a_q(v)=\langle d(h, v), pre_H(v)\rangle$, where $pre_H(v)$ is a flag indicating whether any shortest path from $h$ to $v$ passes through another hub $h'$ ($h' \neq h$ and $h' \neq v$). Quegel starts processing $\langle h\rangle$ by calling {\em init\_activate}() to activate $h$. The UDF $v.${\em init\_value}($\langle h\rangle$) is specified to set $pre_H(v)\gets$ {\em FALSE}, and to set $d(s, v)\gets 0$ if $v=h$ or set $d(s, v)\gets \infty$ otherwise.

The UDF $v.${\em compute}(.) is implemented as follows. In this algorithm, a message sent by $v$ indicates whether there exists a shortest path from $h$ to $v$ that contains another hub $h'\neq h$ (here, $h'$ can be $v$); if so, for any vertex $u\notin H$ newly activated by that message, it holds that $h\not\in H_u$. Based on this idea, the algorithm is given as follows. In superstep~1, $h$ broadcasts message $\langle${\em FALSE}$\rangle$ to its neighbors. In superstep~$i$ ($i>1$), if $d(h, v)\neq\infty$, then $v$ is already visited by BFS, and it votes to halt directly; otherwise, $v$ is activated for the first time, and it sets $d(h, v) \gets step_q-1$, and receives and processes incoming messages as follows. If $v$ receives $\langle${\em TRUE}$\rangle$ from a neighbor $w$, then a shortest path from $h$ to $v$ via $w$ passes through another hub $h'$ ($h' \neq h$ and $h' \neq v$), and thus $v$ sets $pre_H(v)\gets$ {\em TRUE}. Then, if $v\in H$ or $pre_H(v)=$ {\em TRUE}, $v$ broadcasts message $\langle${\em TRUE}$\rangle$ to each neighbor $u$; otherwise, $v$ broadcasts message $\langle${\em FALSE}$\rangle$ to all its neighbors. Finally, $v$ votes to halt.

To compute $L(v)$ using the above algorithm, we specify the query-independent attribute of a vertex $v$ as $a^V(v)=(\Gamma(v), L(v))$, where $L(v)$ is initially empty. After a query $\langle h\rangle$ is processed, we perform the following operation in the query dumping UDF: (i)~if $v\notin H$, $v$ adds $\langle h, d(h,v)\rangle$ to $L(v)$ only if $pre_H(v)=$ {\em FALSE}; (ii)~if $v\in H$, $v$ always adds $\langle h, d(h,v)\rangle$ to $L(v)$.

After all the $|H|$ queries are processed, $L(v)$ is fully computed for each $v\in V$. Then, each vertex $v$ saves $L(v)$ along with other V-data to HDFS, which is to be loaded later by the Quegel program for processing PPSP queries described previously.

\vspace{2mm}

\noindent{\bf Extension to Directed Graphs.} If $G$ is directed, we make the following changes. First, each vertex $v$ now has in-degree $|\Gamma_{in}(v)|$ and out-degree $|\Gamma_{out}(v)|$, and thus we consider three different ways of picking hubs, i.e., picking those vertices with the highest (i)~in-degree, or (ii)~out-degree, or (iii)~sum of in-degree and out-degree. Second, each vertex $v$ now maintains two core-hub sets: an entry-hub set $H^{in}_v$ and an exit-hub set $H^{out}_v$. A hub $h\in H$ is an entry-hub (exit-hub) of $v$, iff no other hub $h'$ ($\neq h,v$) exists on any shortest path from $v$ to $h$ (from $h$ to $v$). Accordingly, we obtain two lists of hub-distance labels, $L_{in}(v)$ and $L_{out}(v)$. During indexing, we construct $L_{in}(v)$ ($L_{out}(v)$) by backward (forward) BFS, i.e., sending messages to in-neighbors (out-neighbors). When answering PPSP queries, we compute $d_{ub}$ similarly but $h_s\in H_s$ (and $h_t\in H_t$) is now replaced by $h_s\in H^{in}_s$ (and $h_t\in H^{out}_t$).

\subsection{XML Keyword Search}\label{ssec:xml}

Section~\ref{ssec:ppsp} illustrated how graph indexing itself can be formulated as an individual Quegel program. We now present another application of Quegel, i.e., keyword search on XML documents, which makes use of the distributed indexing interface of Quegel described in Section~\ref{sec:interface} directly. Compared with traditional algorithms that rely on disk-based indexes~\cite{LiuC08pvldb,ZhouBWLCLG12icde}, our Quegel algorithms are much easier to program, and they avoid the expensive cost of constructing any disk-based index. Although simple MapReduce solution has also been developed, it takes around 15 seconds to process each keyword query on an XML document whose size is merely 200MB~\cite{ZhangMWZ10dasfaa}. The low efficiency is because MapReduce is not designed for querying workload. In contrast, our Quegel program answers the same kind of keyword queries on much larger XML documents in less than a second. Let us first review the query semantics of XML keyword search, and then discuss XML keyword query processing in Quegel, followed by applications of the query in an online shopping platform.

\vspace{2mm}

\begin{figure}[!t]
    \centering
    \includegraphics[width=\columnwidth]{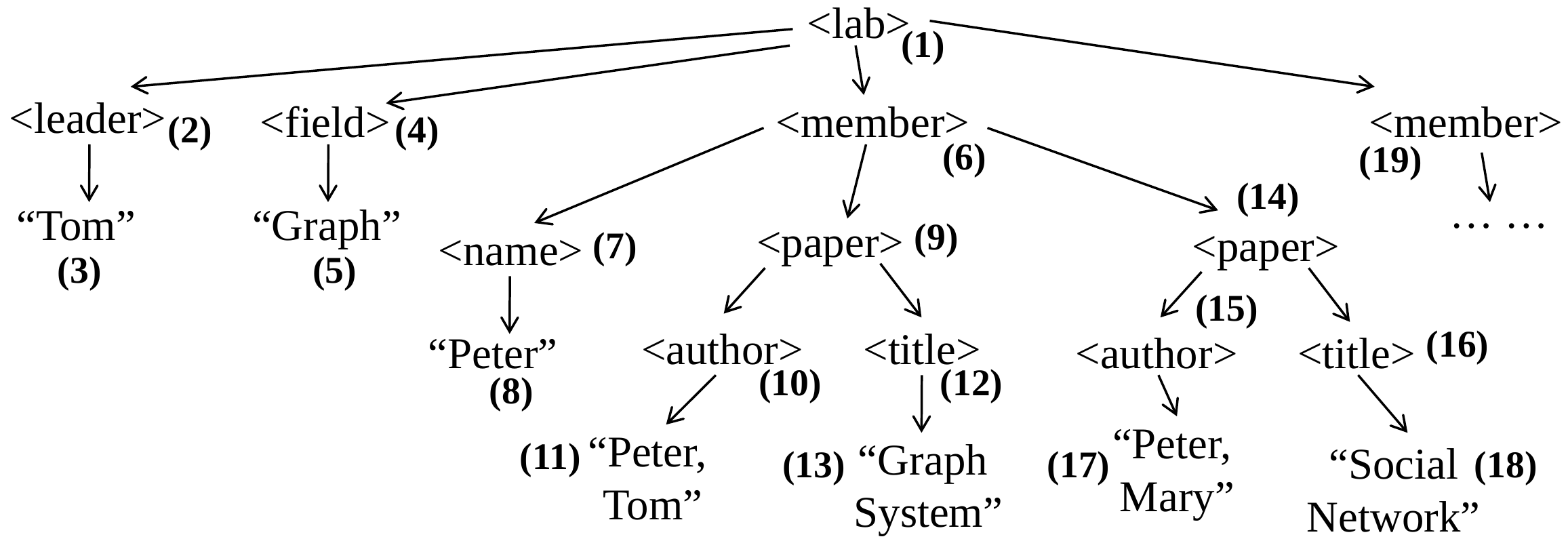}
    \caption{A fragment of an XML document}\label{xml}
\end{figure}

\subsubsection{Query Semantics}
An XML document can be regarded as a rooted tree, where internal vertices are XML tags and leaf vertices are texts. To illustrate, Figure~\ref{xml} shows the tree structure of an XML document describing the information of a research lab. We denote the set of words contained in the tag or text of a vertex $v$ by $\psi(v)$, and if a keyword $k\in \psi(v)$, we call $v$ as a \emph{matching vertex} of $k$ (or, $v$ matches $k$). Given an XML document modeled by a tree $T$, an XML keyword query $q=\{k_1$, $k_2$, $\ldots$, $k_m\}$ finds a set of trees, each of which is a fragment of $T$, denoted by $R$, such that for each keyword $k_i\in q$, there exists a vertex $v$ in $R$ matching $k_i$. We call each result tree $R$ as a {\em matching tree} of $q$.

Different semantics have been proposed to define what a meaningful matching tree $R$ could be. Most semantics require that the root of $R$ be the {\em Lowest Common Ancestor} (LCA) of $m$ vertices $v_1$, $\ldots$, $v_m$, where each vertex $v_i$ matches a keyword $k_i\in q$. For example, given the XML tree in Figure~\ref{xml} and a query $q=\{$Tom, Graph$\}$, vertex~9 is the LCA of the matching vertices~11 and~13, while vertex~1 is the LCA of the matching vertices~3 and~5.

We consider two popular semantics for the root of $R$: {\em Smallest LCA} (SLCA) and {\em Exclusive LCA} (ELCA)~\cite{ZhouBWLCLG12icde}. For simplicity, we use ``LCA/SLCA/ELCA of $q$'' to denote ``LCA/SLCA/ELCA of matching vertices $v_1$, $\ldots$, $v_m$''. An SLCA of $q$ is defined as an LCA of $q$ that is not an ancestor of any other LCA of $q$. For example, in Figure~\ref{xml}, vertex~9 is the SLCA of $q=\{$Tom, Graph$\}$, while vertex~1 is not since it is an ancestor of another LCA, i.e.\ vertex~9. Let us denote the subtree of $T$ rooted at vertex $v$ by $T_v$, then a vertex $v$ is an ELCA of $q$ if $T_v$ contains at least one occurrence of all keywords in $q$, after pruning any subtree $T_u$ (where $u$ is a child of $v$) which already contains all keywords in $q$. Referring to Figure~\ref{xml} again, both vertices~1 and~9 are ELCAs of $q=\{$Tom, Graph$\}$. Vertex~1 is an ELCA since after pruning the subtree rooted at vertex~$6$, there still exist vertices~3 and~5 matching the keywords in $q$. In contrast, if $q=\{$Peter, Graph$\}$, then vertex~9 is an ELCA of $q$, while vertex~1 is not an ELCA of $q$ since after pruning the subtree rooted at vertex~$6$, there is no vertex matching ``Peter''.

Once the root, $r$, of a matching tree is determined, we may return the whole subtree $T_r$ as the result tree $R$. However, if $r$ is at a top level of the input XML tree, $T_r$ can be large (e.g., the subtree rooted at vertex~1) and may contain much irrelevant information. For an SLCA $r$, {\em MaxMatch}~\cite{LiuC08pvldb} was proposed to prune irrelevant parts from $T_r$ to form $R$. Let $K(v)$ be the set of keywords matched by the vertices in $T_v$. If a vertex $v_1$ has a sibling $v_2$, where $K(v_1) \subset K(v_2)$, then $T_{v_1}$ is pruned. For example, let $q=\{$Tom, Graph$\}$ and consider the subtree rooted at vertex~1 in Figure~\ref{xml}. Since vertex~9 contains \{Tom, Graph\} in its subtree while its sibling vertex~14 does not contain any keyword in its subtree, the subtree rooted at vertex~14 is pruned.

\subsubsection{Query Algorithms}
We now present the Quegel algorithms for computing SLCA, ELCA and MaxMatch. The Quegel program first loads the graph that represents the XML document (the graph is obtained by parsing the XML document with a SAX parser), where each vertex $v$ is associated with its parent $pa(v)$ and its children $\Gamma_{c}(v)$ (V-data). Then, each worker constructs an inverted index from the loaded vertices using the indexing interface described in Section~\ref{sec:interface}.

To process a query $q$, the UDF {\em init\_activate}() activates only those vertices $v$ with $\psi(v)\cap q\neq\emptyset$. The query-independent attribute of each vertex $v$, $a^V(v)$, maintains $pa(v)$, $\Gamma_{c}(v)$, and $\psi(v)$, and the query-dependent attribute $a_q(v)$ maintains a bitmap $bm(v)$, where bit~$i$ (denoted by $bm(v)[i]$) equals 1 if keyword $k_i$ exists in subtree $T_v$ and 0 otherwise. The UDF $v.${\em init\_value}($q$) sets each bit $bm(v)[i]$ to $1$ if $k_i\in\psi(v)$ and 0 otherwise. For simplicity, if all the bits of $bm(v)$ are 1, we call $bm(v)$ as \emph{all-one}. We now describe the query processing logic of $v.${\em compute}(.) for SLCA, ELCA and MaxMatch semantics as follows.

\vspace{2mm}

\noindent{\bf Computing SLCA in Quegel.} In superstep~1, all matching vertices have been activated by {\em init\_activate}(), and each matching vertex $v$ sends $bm(v)$ to its parent $pa(v)$ and votes to halt. In superstep~$i$ ($i>1$), there are two cases in processing a vertex $v$. {\bf Case~(a)}: if some bit of $bm(v)$ is 0, $v$ computes the bitwise-OR of $bm(v)$ and those bitmaps received from its children, which is denoted by $bm_{OR}$. If $bm_{OR}\neq bm(v)$, then some new bit of $bm(v)$ should be set due to a newly matched keyword; thus, $v$ sets $bm(v)=bm_{OR}$, and sends the updated $bm_{v}$ to its parent $pa(v)$. In addition, if $bm_{OR}$ is all-one, then (1)~if $v$ receives an all-one bitmap from a child, $v$ is labeled as a non-SLCA (the label is also maintained in $a_q(v)$); (2)~otherwise, $v$ is labeled as an SLCA. {\bf Case~(b)}: if $bm(v)$ is all-one, then $v$ has been labeled either as an SLCA or as a non-SLCA (because a descendant is an SLCA) in an earlier superstep. (1)~If $v$ is labeled as a non-SLCA, $v$ votes to halt directly; while (2)~if $v$ is labeled as an SLCA, and $v$ receives an all-one bitmap from a child, then $v$ labels itself as a non-SLCA. Finally, $v$ votes to halt.

In the above algorithm, a vertex may send messages to its parent multiple times. To make sure that each vertex sends at most one message to its parent, we design another {\bf level-aligned} algorithm as follows. Specifically, we pre-compute the level of each vertex $v$ in the XML tree, denoted by $\ell(v)$, by performing BFS from the tree root (with a traditional Pregel job). Then, our Quegel program loads the preprocessed data, where each vertex $v$ also maintains $\ell(v)$ in $a_q(v)$. The UDF $v.${\em compute}(.) is designed as follows. Initially, we use an aggregator to collect the maximum level of all the matching vertices, denoted by $\ell_{max}$. The aggregator maintains $\ell_{max}$ and decrements it by one after each superstep. In a superstep, a vertex $v$ at level $\ell_{max}$ computes the bitwise-OR of $bm(v)$ and all the bitmaps received from its children at level $(\ell_{max}+1)$; the bitwise-OR is then assigned to $bm(v)$ and sent to $v$'s parent $pa(v)$. Moreover, if an all-one bitmap is received, $v$ labels itself as a non-SLCA directly; otherwise, and if $bm(v)$ becomes all-one, then $v$ labels itself as an SLCA. Finally, $v$ votes to halt. Note that those matching vertices $u$ with $\ell(u)<\ell_{max}$ remain active until they are processed.

\vspace{2mm}

\noindent{\bf Computing ELCA in Quegel.} We use a level-aligned algorithm to compute ELCAs as follows. In a superstep, an active vertex $v$ at level $\ell_{max}$ updates $bm(v)$ and sends it to the parent $pa(v)$ as in SLCA computation. Meanwhile, $v$ also computes another bitmap $bm^*_{OR}$ (in addition to $bm_{OR}$), which is the bitwise-OR of $bm(v)$ (before its update) and all the non-all-one bitmaps received from its children at level $(\ell_{max}+1)$. And $v$ labels itself as an ELCA if $bm^*_{OR}$ is all-one.

In our SLCA and ELCA algorithms, each vertex $v$ also maintains in $a^V(v)$ its start and end positions in the XML document, denoted by $start(v)$ and $end(v)$, which are also obtained during the SAX parsing. After a query is processed, each vertex that is labeled as an SLCA or ELCA dumps $[start(v), end(v)]$ to HDFS, so that users can obtain $T_v$ by reading the corresponding part of the XML document.

\vspace{2mm}

\noindent{\bf Computing MaxMatch in Quegel.}  Our Quegel algorithm for computing MaxMatch prunes irrelevant parts from the subtree rooted at each SLCA, and all vertices in the result matching trees dump themselves to HDFS after a query is processed, which can then be sent to the client and assembled as trees for display.

The algorithm is also level-aligned, and consists of two phases. In Phase~1, we run a variant of the level-aligned SLCA algorithm, where each vertex $v$ sends message $\langle v, bm(v)\rangle$ to $pa(v)$. When a vertex $v$ receives a message $\langle u, bm(u)\rangle$ from a child $u$, it keeps $\langle u, bm(u)\rangle$ in $a_q(v)$. To avoid the algorithm from terminating after Phase~1, we keep the SLCA vertices active (i.e., they do not vote to halt) during the computation of Phase~1. Phase~1 ends when the superstep that decrements $\ell_{max}$ to $0$ finishes, and then the aggregator sets the phase number as 2 to start Phase~2.

Phase~2 performs downward propagation from those SLCAs found by Phase~1. In a superstep, each active vertex $v$ labels itself to indicate that $v$ is in a matching tree $R$ (the label is also kept in $a_q(v)$). Then, $v$ sends messages to its children that are not dominated by any of their siblings. Here,  $u_1$ is dominated by  $u_2$ if $K(u_1)\subset K(u_2)$, and we check the condition using their bitmaps as follows: $bm(u_1)\neq bm(u_2)$ and $(bm(u_1)$ {\em Bit-OR} $bm(u_2))=bm(u_2)$. In this way, dominated subtrees are pruned from $R$, and Quegel dumps only the labeled vertices to HDFS.

\vspace{2mm}

\noindent{\bf Applications of XML Keyword Search.} Though originally proposed for querying a single XML document~\cite{LiuC08pvldb,ZhouBWLCLG12icde}, our algorithms can also be used to query a large corpus of many XML documents. We illustrate this by one application in e-commerce. During online shopping, a customer may pose a keyword query (in the form of an AJAX request) from a web browser to search for interested products. The web server will obtain the matched products from the database, organize them as an XML document, and send it back to the client side. The browser of the client will then parse the XML document by a Javascript script to display the results. The server may log the various AJAX responses to disk, so that data scientists and sellers may pose XML keyword queries on the logged XML corpus to study customers' search behaviors of specific products, to help them make better business decisions.

\subsection{Terrain Shortest Path Queries}\label{ssec:terrain}
Technological advance in remote sensing has made available high resolution terrain data of the entire Earth surface. Terrain data are usually represented in the {\em Digital Elevation Model} (DEM), which is an elevation mesh of sampled ground positions at regularly spaced interval. Since terrain data are usually collected at high resolution (e.g., 10m sampling intervals), the data size is usually huge.

\begin{figure}[!t]
\centering
\subfigure[TIN]{\includegraphics[width=0.49\columnwidth]{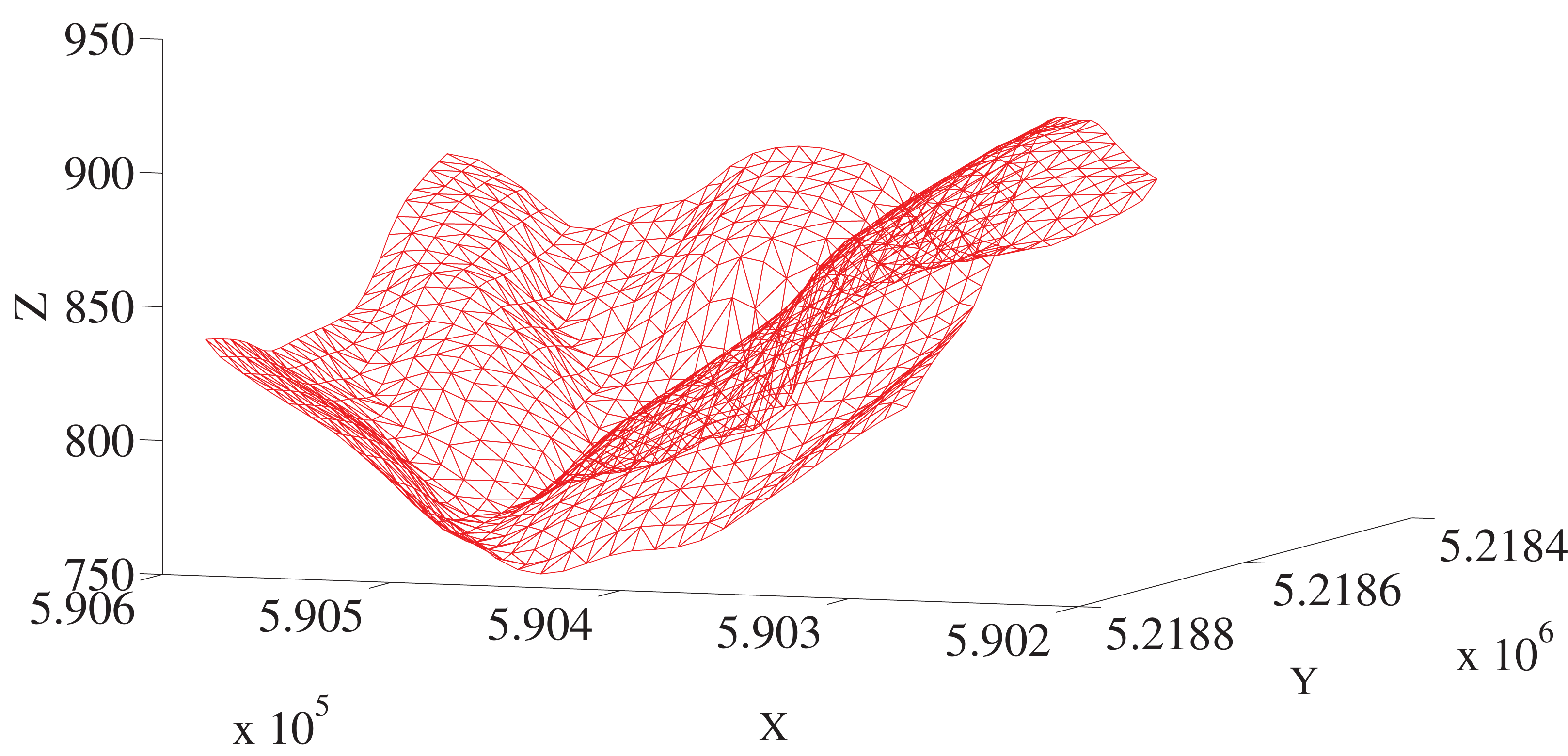}\label{tin}}
\subfigure[Network Distance]{\includegraphics[width=0.49\columnwidth]{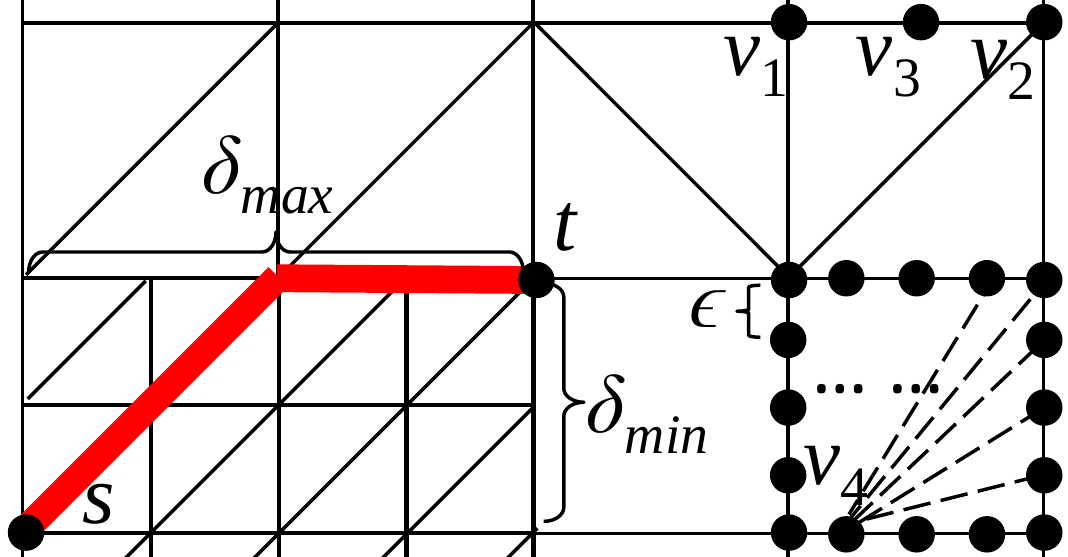}\label{net}}
\caption{Terrain data model}
\end{figure}

Many recent studies propose algorithms for processing various spatial queries over terrain data, including P2P shortest path queries~\cite{LiuW11sigmod}, nearest neighbor (NN) queries~\cite{ShahabiTX08pvldb,KaulWYJ13pvldb} and reverse NN queries~\cite{YanZN12cikm1,KaulWYJ13pvldb}. Applications of terrain queries include disaster response, outdoor activities, and military operations~\cite{YanZN12cikm1}. Existing works adopt the {\em Triangulated Irregular Network} (\emph{TIN}) terrain representation as illustrated in Figure~\ref{tin}, which is derived from the DEM data. Since the terrain surface is composed of triangular faces, existing works use {\em Chen and Han's algorithm}~\cite{KanevaO00cccg}, which is a polyhedron shortest path algorithm, to compute the terrain shortest path between two terrain locations. This approach has very poor performance and scalability, since the time complexity of Chen and Han's algorithm is quadratic to the number of triangular faces. For example, even with surface simplification (with precision loss), the algorithm of~\cite{LiuW11sigmod} can only process terrain shortest paths with length of merely several hundred meters, and it takes hundreds to thousands of seconds to compute one such shortest path. We propose an efficient approximate solution with a much lower cost.

Let $d_N(u, v)$ be the network (which is TIN here) distance between two vertices (i.e., locations), $u$ and $v$. Here, $d_N(u, v)$ upper-bounds the actual terrain distance between $u$ and $v$, since the TIN shortest path is also a path on the terrain surface. However, the TIN shortest path can be very different from the actual terrain shortest path~\cite{ShahabiTX08pvldb}. We further show that the difference cannot be effectively reduced simply by increasing the sampling rate. Consider the mesh fragment shown in Figure~\ref{net}, and suppose that all vertices have the same elevation. If only horizontal and vertical edges are considered, then no matter what the sampling interval is, $d_N(s, t)$ is lower bounded by the Manhattan distance between $s$ and $t$, even though the terrain shortest path is given by a straight line between $s$ and $t$. Now consider a TIN where faces are diagonally triangulated, we can show that $d_N(s, t)$ is lower bounded by $\delta_{max}+(\sqrt{2}-1)\cdot\delta_{min}$, where $\delta_{max}$ (and $\delta_{min}$) refers to the larger (and smaller) one of $|s.x-t.x|$ and $|s.y-t.y|$. Thus, a better solution is needed.

The above discussion motivates us to propose a new transformation from the terrain data to a network that gives more accurate terrain shortest path distance, and we can use Quegel to achieve efficient computation on the network. The idea is to add shortcut edges as illustrated by the last grid cell in Figure~\ref{net}. Specifically, we split each edge of a cell by adding vertices such that the distance between two neighboring vertices is no more than $\epsilon$ as shown in Figure~\ref{net}. Then, in each cell, we add a straight line between every pair of vertices that are not on the same horizontal or vertical edge. We then compute the shortest path on the new network to approximate the terrain shortest path. Since the cell shortcuts are in different directions, the network shortest path can be close to the actual terrain shortest path. Note that even TIN just interpolates the elevation of an arbitrary location from the sampled elevation data~\cite{LiuW11sigmod}, as the actual elevation is not known. For example, in Figure~\ref{net}, the elevation of $v_3$ is linearly interpolated from samples $v_1$ and $v_2$. Therefore, the shortest paths computed on  TIN~\cite{LiuW11sigmod} and our graph model both just approximates the actual shortest path.

Now the problem of computing P2P shortest path over the terrain is transformed to the problem of computing the P2P shortest path in the transformed network. Since terrain data can be of planetary scale and the transformed network is even larger, we employ Quegel for distributed shortest path querying in the transformed network. The logic of the {\em compute}(.) function can simply be the distributed single-source shortest-path (SSSP) algorithm of~\cite{pregel}, where each active vertex updates its current distance (from $s$) using that of its neighbors, and propagates the updated distance to its neighbors to activate them for further distance computation, until the process converges. We further devise a mechanism to terminate the SSSP computation earlier (without traversing all the vertices) when $s$ and $t$ are close to each other, which is described as follows.

Let $d_E(u, v)$ be the Euclidean distance between vertices $u$ and $v$. In any superstep, when a vertex $v$ is currently active (i.e., $v$ is at the distance propagation wavefront), $d_N(s, v)$ is updated based on $d_N(s, u)$ sent by the neighbors $u$ of $v$ from the previous superstep. Meanwhile, we compute $d_E(s, v)$ using the coordinates of $s$ and $v$. Note that $d_E(s, v)$ lower-bounds $d_N(s, v)$. We use the aggregator to compute the minimum value of $d_E(s, v)$, denoted by $d_E^{min}$, among all vertices $v$ at the distance propagation wavefront. If $d_N(s, t)<d_E^{min}$, vertex~$t$ calls {\em force\_terminate}() to end the computation. This is because for any vertex $w$ that will be at the distance propagation wavefront in any following superstep, we have $d_N(s, w)>d_E^{min}$ for the current $d_E^{min}$. However, we already have $d_N(s, t)<d_E^{min}$, and thus no $d_N(s, w)$ computed in any following superstep (including $d_N(s, t)$) can be smaller than the current $d_N(s, t)$.

In our actual implementation, to avoid a large number of supersteps caused by large graph diameter, we use the idea of~\cite{blogel} which first partitions the graph into subgraphs that group spatially close vertices, and then propagates the distance updates from $s$ in the unit of subgraphs (instead of vertices). Experiments in Section~\ref{sec:results} verify that our new method computes high-quality terrain shortest paths very efficiently for any path length (in contrast to only several hundred meters as in~\cite{LiuW11sigmod}).

\subsection{P2P Reachability}\label{ssec:reach}
In this section, we consider P2P reachability query $(s, t)$, which determines whether there exists a path from $s$ to $t$ in a directed graph $G=(V, E)$. The Quegel algorithms for BFS and BiBFS as described in Section~\ref{sssec:no_idx} are also applicable to this problem. We now consider the Quegel solution that makes use of indexing.

A P2P reachability query on a direct graph $G$ can be reduced to one on a directed acyclic graph (DAG) $G'$. Each vertex of $G'$ represents a strongly connected component (SCC) of $G$, and each edge represents the fact that one component can reach another. To answer whether $u$ can reach $v$ in $G$, we simply look up their corresponding SCCs, $S_u$ and $S_v$, respectively, which are the nodes in $G'$. Vertex $u$ can reach $v$ in $G$ iff $S_u=S_v$ or $S_u$ can reach $S_v$ in $G'$. Note that the SCCs of $G$ can be computed in Pregel using the algorithm of~\cite{pregelplus}, which associates each vertex $v$ in $G$ with its corresponding (SCC) vertex $S_v$ in $G'$. This $v$-to-$S_v$ mapping relation can be pre-computed as an independent Pregel job, and stored on HDFS to be loaded later by Quegel workers into their local index field {\em Worker::index}. When a query $(s,t)$ arrives, each worker may look up $S_s$ and $S_t$ from the index and activate them (if they reside in the worker) in {\em init\_activate}(). For ease of discussion, we assume $G$ is a DAG hereafter.

Existing work on P2P reachability indexing combines graph traversal with vertex-label based pruning in order to be scalable, such as~\cite{ZhangYQZZ12edbt,WeiYLJ14pvldb}. Due to the requirement of graph traversal, the graph and the vertex labels have to reside in main memory, and for massive graphs, one has to resort to a distributed main-memory system. In this section, we demonstrate how the index of~\cite{ZhangYQZZ12edbt} can be used in Quegel. We assume that a depth-first search forest of $G$ is given (which is required by the no-label to be introduced shortly), so that each vertex $v$ knows its parent $pa(v)$ in the forest, the pre-order number $pre(v)$ and the post-order number $post(v)$. This can be computed in memory, or using the IO-efficient algorithm of~\cite{IODFS}.

During the indexing phase, we compute three labels for each vertex $v$ using three cascaded Pregel jobs: (1)~level $\ell(v)$, (2)~yes-label $yes(v)$ and (3)~no-label $no(v)$. These labels are then used in our Quegel algorithm to prune vertices from further expanding during bidirectional BFS from $s$ and $t$.

\vspace{2mm}

\begin{figure}[t]
    \centering
    \includegraphics[width=0.7\columnwidth]{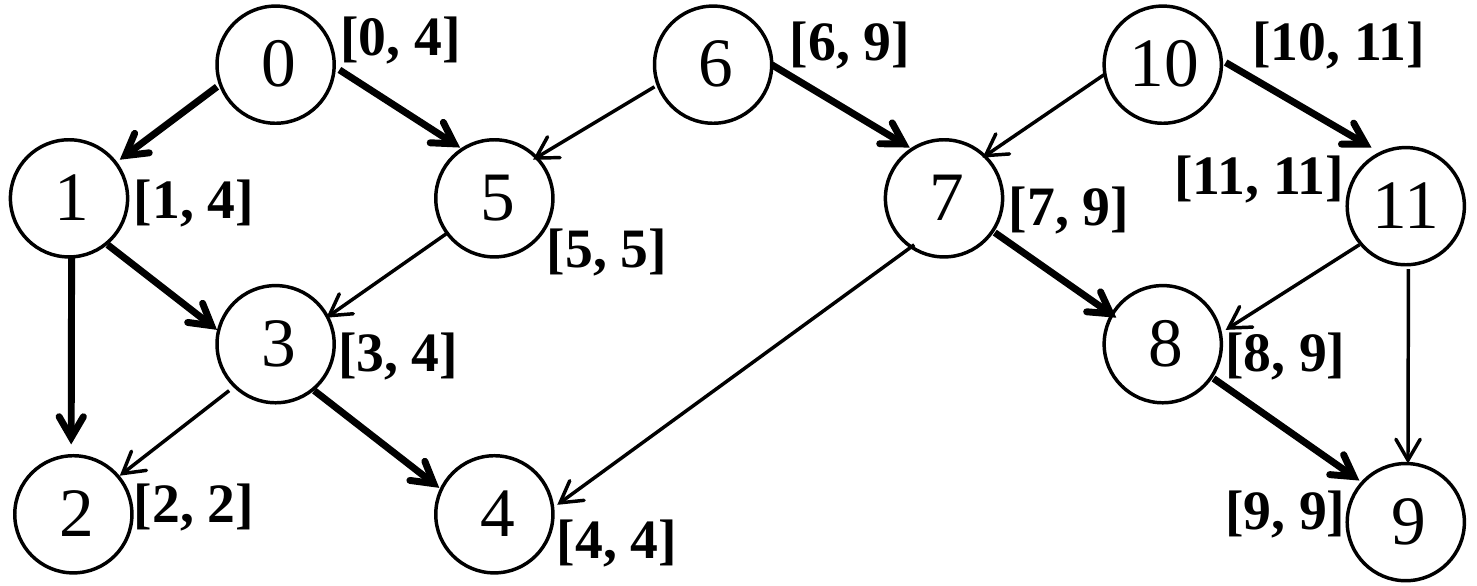}
    \caption{Illustration of yes-labels}\label{yesLabel}
\end{figure}

\noindent{\bf Level Label.} We first define $\ell(v)$ and discuss its computation. Let us call a vertex with zero in-degree as a root, then $\ell(v)$ is defined as the largest number of hops from a root to $v$. For example, consider the DAG shown in Figure~\ref{yesLabel}. Vertex~9 has level 3 although it is just two hops away from root~10 (through path $10\rightarrow 11\rightarrow 9$), since the longest path from root~10 has three hops (e.g., path $10\rightarrow 7\rightarrow 8\rightarrow 9$).

According to the definition of the level label, if $u$ can reach $v$, then $\ell(u)<\ell(v)$. Therefore, in our Quegel algorithm, if the forward BFS from $s$ activates a vertex $v$ with $\ell(v)\geq\ell(t)$, $v$ votes to halt directly as it cannot reach $t$; similarly, if the backward BFS from $t$ activates a vertex $v$ with $\ell(s)\geq\ell(v)$, $v$ votes to halt directly as $s$ cannot reach $v$. Note that the vertex labels of $s$ and $t$ can be obtained using aggregator at the beginning of a query $(s, t)$, so that any vertex $v$ can get them from the aggregator in {\em compute}(.).

The Pregel algorithm for level computation is as follows. Initially, only roots $r$ are active with $\ell(r)=0$, while $\ell(v)$ is initialized as $\infty$ for all other vertices $v$. In superstep~1, each root $r$ broadcasts $\ell(r)$ to its out-neighbors before voting to halt. In superstep~$i$ ($i>1$), each active vertex $v$ gets the largest incoming message $\ell(u)$ (sent from in-neighbor $u$); here, we know that $v$'s level should be at least $\ell(u)+1$, and thus we check if $\ell(u)+1>\ell(v)$. If so, $v$ updates $\ell(v)=\ell(u)+1$, and broadcasts $\ell(v)$ to all its out-neighbors. Finally, $v$ votes to halt. Upon convergence, for each vertex $v$, $\ell(v)$ equals the level of $v$.

\vspace{2mm}

\noindent{\bf Yes-Label.} We now define $yes(v)$ and discuss its computation. Recall that the pre-order number $pre(v)$ is available for each vertex $v$. Let us define $Out(v)$ to be the set of all vertices reachable from $v$ (including $v$ itself), then $yes(v)$ is defined as $[pre(v), \max_{u\in Out(v)}pre(u)]$. As an illustration, consider the graph shown in Figure~\ref{yesLabel}, where the bold edges belong to the DFS forest, and the vertices are marked with their pre-order numbers. Vertex~5 has yes-label $[5, 5]$ since the largest vertex that it can reach is itself; while vertex~7 has yes-label $[7, 9]$ as the largest vertex that it can reach has ID 9.

The yes-label has the following property: if $yes(v)\subseteq yes(u)$, then $u$ can reach $v$~\cite{ZhangYQZZ12edbt}. To illustrate, in Figure~\ref{yesLabel}, we can conclude that vertex~0 can reach vertex~2 since $[2, 2]\subseteq[0, 4]$. In fact, this property holds as long as $pre(v)$ is computed from a spanning forest of $G$ (including a DFS forest). Intuitively, $yes(v)\subseteq yes(u)$ iff $u$ is an ancestor of $v$ in the forest. Therefore, in our Quegel algorithm, if the forward BFS from $s$ activates a vertex $v$ with $yes(t)\subseteq yes(v)$, $v$ calls {\em force\_terminate}() and indicates that $s$ can reach $t$. This is because $v$ is obviously reachable from $s$, and $v$ can reach $t$ according to the yes-labels. Similarly, if the backward BFS from $t$ activates a vertex $v$ with $yes(v)\subseteq yes(s)$, $v$ calls {\em force\_terminate}() and indicates that $s$ can reach $t$.

To compute the yes-labels, we only need to compute $max(v)=\max_{u\in Out(v)}pre(u)$ for each vertex $v$ as follows. Initially, for each vertex $v$, $max(v)$ is initialized as $pre(v)$, and only those vertices with zero out-degree are active; each active vertex $v$ sends $max(v)$ to its in-neighbors in superstep~1 and votes to halt. In superstep~$i$ ($i>1$), each vertex $v$ receives the incoming messages, and let the largest one be $max(u)$; if $max(u)>max(v)$, $v$ sets $max(v)=max(u)$ and broadcasts $max(v)$ to its in-neighbors; finally, $v$ votes to halt.

A weakness of this algorithm is that, a vertex $v$ may update $max(v)$ and broadcast $max(v)$ to its in-neighbors for more than once. We design a more efficient level-aligned algorithm that makes use of level $\ell(v)$ to ensure that each vertex $v$ only updates and broadcasts $max(v)$ once, which works as follows. Initially, only those vertices with zero out-degree are active, and we use aggregator to collect their maximum level $\ell_{max}$. Then, the aggregator maintains $\ell_{max}$ and decrements it by one after each superstep; all vertices $v$ with $\ell(v)>\ell_{max}$ are already processed, while all vertices $u$ with $\ell(u)=\ell_{max}$ are being processed. In a superstep, a vertex $v$ receives messages, and let the largest one be $max(u)$; if $max(u)>max(v)$, $v$ sets $max(v)\gets max(u)$. Then, each vertex $v$ with $\ell(v)=\ell_{max}$ broadcasts $max(v)$ to its in-neighbors and votes to halt.

\begin{figure}[t]
    \centering
    \includegraphics[width=0.7\columnwidth]{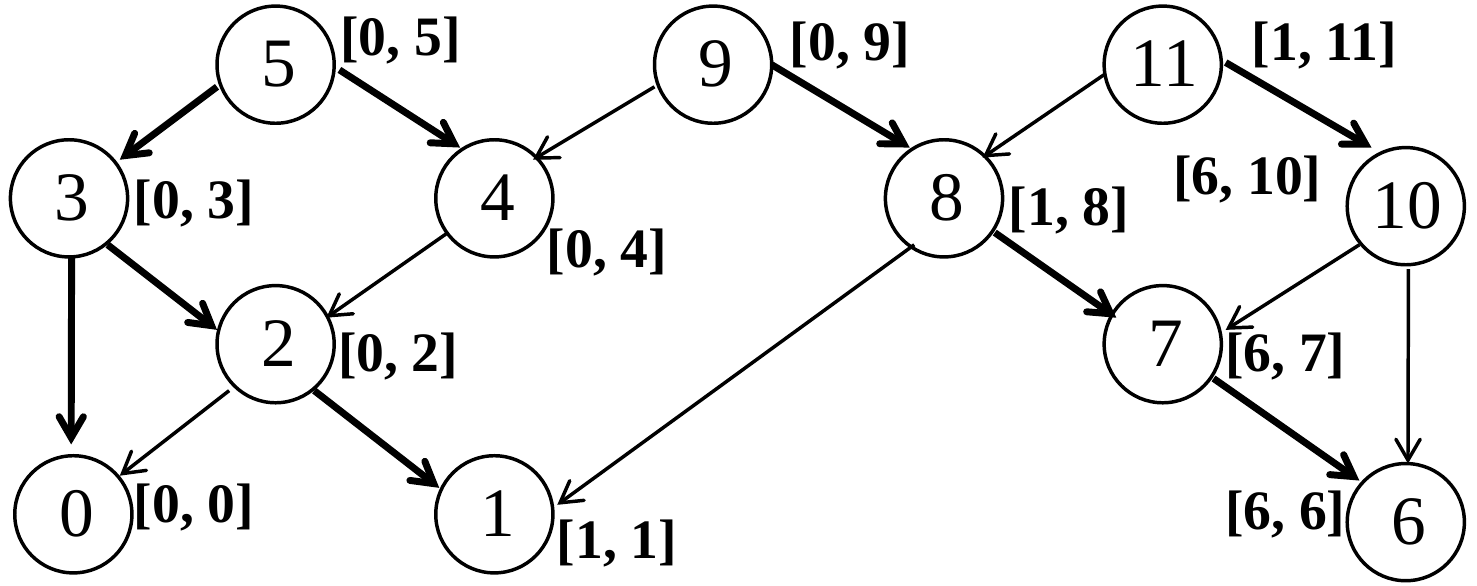}
    \caption{Illustration of no-labels}\label{noLabel}
\end{figure}

\vspace{2mm}

\noindent{\bf No-Label.} Finally, we define $no(v)$ and discuss its computation. For each vertex $v$, $no(v)$ is defined as $[\min_{u\in Out(v)}post(u), post(v)]$. As an illustration, consider the graph shown in Figure~\ref{noLabel}, which is the same one as in Figure~\ref{yesLabel} except that the vertices are marked with their post-order numbers. Vertex~4 has no-label $[0, 4]$ since the smallest vertex that it can reach has ID 0; while vertex~8 has no-label $[1, 8]$ as the smallest vertex that it can reach has ID 1.

The no-label has the following property: if $u$ can reach $v$, then $no(v)\subseteq no(u)$~\cite{ZhangYQZZ12edbt}. The property can be easily observed from Figure~\ref{noLabel}. We actually use its contrapositive: if $no(v)\not\subseteq no(u)$, then $u$ cannot reach $v$. To illustrate, in Figure~\ref{noLabel}, we can conclude that vertex~11 cannot reach vertex~0 since $[0, 0]\not\subseteq[1, 11]$. Therefore, in our Quegel algorithm, if the forward BFS from $s$ activates a vertex $v$ with $no(t)\not\subseteq no(v)$, $v$ votes to halt directly as $v$ cannot reach $t$; similarly, if the backward BFS from $t$ activates a vertex $v$ with $no(v)\not\subseteq no(s)$, $v$ votes to halt directly as $s$ cannot reach $v$.

The Pregel algorithm for no-label computation is symmetric to that for yes-label computation, and is thus omitted.

\subsection{Graph Keyword Search}\label{ssec:gkey}
In this section, we consider a simplified version of the graph keyword search problem~\cite{HeWYY07sigmod} which was recently studied by~\cite{QinYCCZL14sigmod} on MapReduce: given a keyword query $Q=\{k_1, k_2, \ldots, k_m\}$ over a graph $G=(V, E)$ where each vertex $v\in V$ has text $\psi(v)$, a keyword search finds a set of rooted trees in the form of $(r,$ $\{\langle v_1, hop(r, v_1)\rangle$, $\langle v_2, hop(r, v_2)\rangle)$, $\ldots$, $\langle v_m, hop(r, v_m)\rangle\})$, where $r$ is the root vertex, and $v_i$ is the closest vertex to $r$ whose text $\psi(v_i)$ contains keyword $k_i$. Moreover, the maximum distance allowed from a root to a matched vertex is constrained to be $\delta_{max}$. Note that a root vertex $r$ determines a unique answer, since we pick the matching vertex closest to $r$ for each keyword.

A simple vertex-centric algorithm for graph keyword search is described as follows. Each vertex $v$ maintains for each $k_i$ a field $\langle v_i, hop(v, v_i)\rangle$ indicating its closest matching vertex $v_i$. Initially, if $k_i\in\psi(v)$, we set $\langle v_i, hop(v, v_i)\rangle=\langle v, 0\rangle$; otherwise, we set $\langle v_i, hop(v, v_i)\rangle=\langle?, \infty\rangle$. Only vertices whose text contains at least one keyword are active. In superstep~1, each matching vertex $v$ finds its fields with $v_i\neq\ ?$ (i.e., $k_i\in\psi(v)$), sends $\langle v_i, hop(v, v_i)\rangle$ to all its in-neighbors, and votes to halt. In superstep~$i$ ($i>1$), a vertex $v$ receives messages $\langle u_i, hop(u, u_i)\rangle$ from its out-neighbors $u$. Here, message $\langle u_i, hop(u, u_i)\rangle$ indicates that vertex $u_i$ matches $k_i$, and it is $hop(u, u_i)$ hops from $u$ (and $u$ is one hop from $v$). Therefore, let $u^*$ be the out-neighbor of $v$ with the smallest $hop(u, u_i)$ and let the matching vertex be $u_i^*$, then if $hop(u^*, u^*_i)+1<hop(v, v_i)$, $v$ updates $\langle v_i, hop(v, v_i)\rangle=\langle u_i, hop(u^*, u^*_i)+1\rangle$ and sends it to all its in-neighbors, before voting to halt. If the computation proceeds after $\delta_{max}$ supersteps, all vertices vote to halt directly and the algorithm stops; by then any vertex $r$ whose $v_i\neq\ ?$ for all keywords $k_i$ corresponds to a result.

\begin{figure}[t]
    \centering
    \includegraphics[width=0.7\columnwidth]{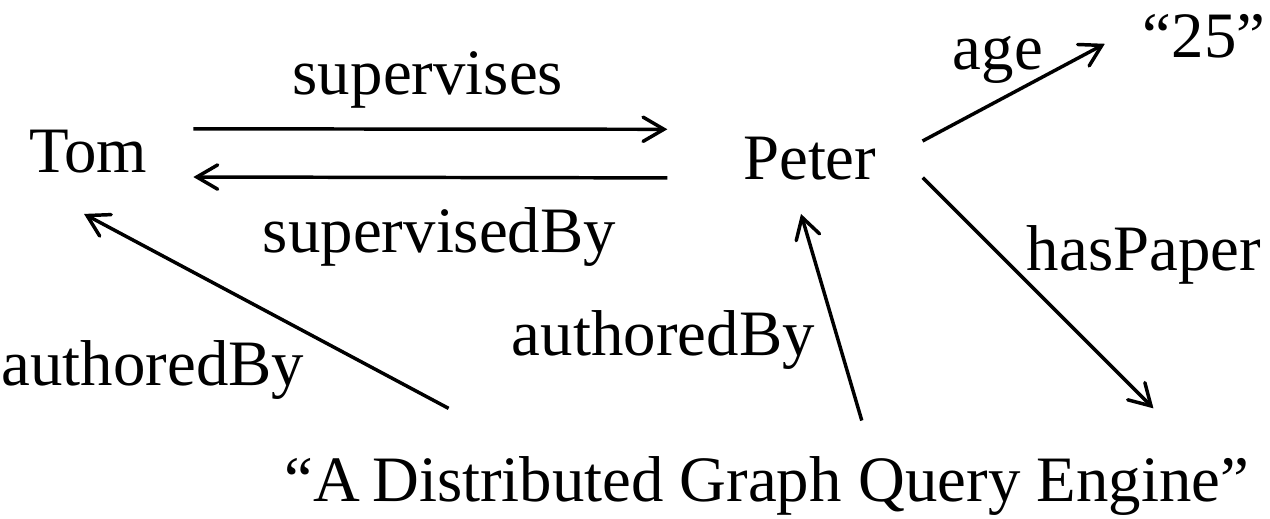}
    \caption{RDF Data Fragment}\label{rdf}
\end{figure}

A typical application of graph keyword search is over RDF data. An RDF data consists of triples of the form $(s, p, o)$ , where $s$, $p$ and $o$ are called as subject, predicate and object, respectively. Conceptually, each triple can be regarded as a directed edge from vertex $s$ to vertex $o$ with edge label $p$, and thus, the whole RDF data can be regarded as a labeled graph. As an illustration, consider the RDF graph shown in Figure~\ref{rdf}, which contains triples like (Tom, supervises, Peter), and (Peter, age, ``25''). Here, the text of some vertex uniquely determines the vertex identity, such as the vertex labeled ``Peter''. The text of such a vertex is called a {\em resource}, which is usually a URI. While for some vertex like ``25'', the text is just a {\em literal} that indicates the value of its predicate, and the text of another vertex can also be this literal.

To perform keyword search over an RDF graph, we first need to convert the set of triples into an adjacency list representation. For a literal vertex $o$ in triple $(s, p, o)$, we store it as an attribute of resource vertex $s$ with attribute $p$ having value $o$. For each resource vertex $v$, two lists are stored, $\Gamma_{in}(v)$ that contains $v$'s in-neighbors (which are resource vertices), and $A(v)$ that contains $v$'s literal out-neighbors. The lists can be easily obtained by MapReduce. For example, to get the in-neighbor lists for all vertices, each mapper splits a triple $(s, p, o)$ (where $o$ is a resource) into $\langle o, (p, s)\rangle$, and each reducer merges all in-neighbors $(s, p)$ of $o$ into $\Gamma_{in}(o)$. Here, each neighbor $s$ is associated with an edge label $p$.

\begin{figure}[t]
    \centering
    \includegraphics[width=\columnwidth]{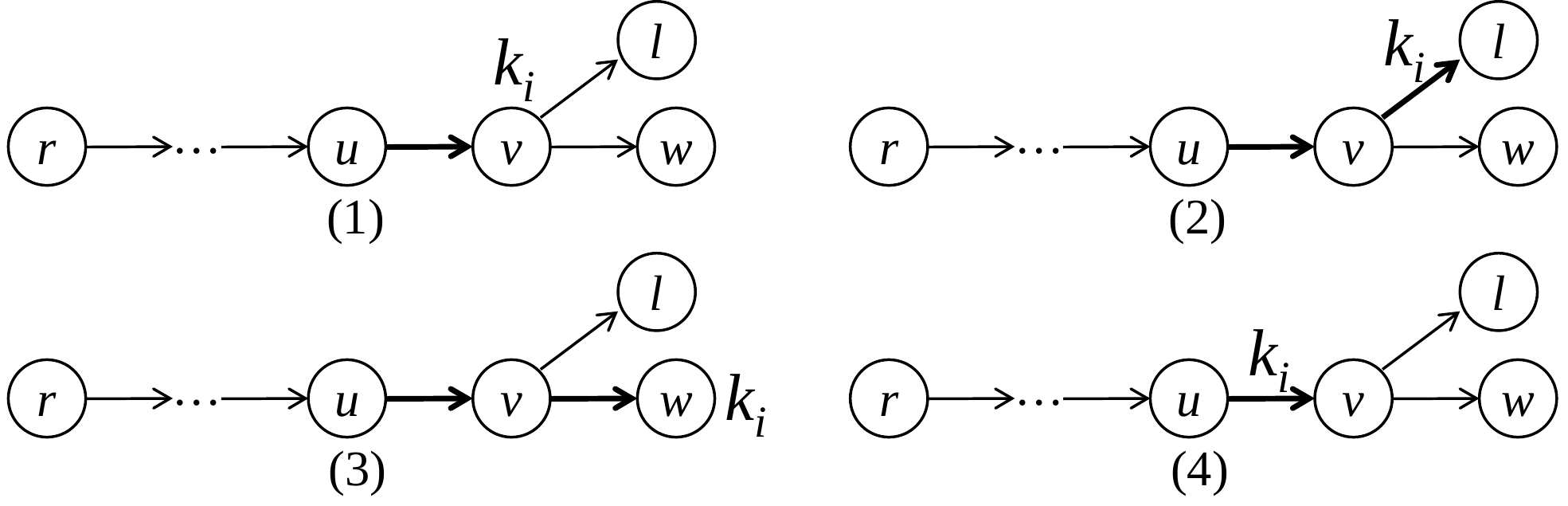}
    \caption{RDF Data Fragment}\label{send}
\end{figure}

The Quegel algorithm for RDF keyword search maintains an inverted index as described in Section~\ref{sec:interface} similar to that for XML keyword search, and only the matching vertices are activated at the beginning of a query. Unlike the vertex-centric algorithm mentioned above, a neighbor $u$ in $\Gamma_{in}(v)$ or $A(v)$ also contains an edge label $p(u)$ which may also match the keywords. Accordingly, we activate a vertex $v$ when any of $\psi(v)$, $\Gamma_{in}(v)$ or $A(v)$ covers a keyword. Moreover, when $v$ sends messages, we need to consider four cases as Figure~\ref{send} illustrates.

We now describe the four cases. Consider a specific keyword $k_i$, (1)~if $k_i\in\psi(v)$, $v$ broadcasts $\langle v, 0\rangle$ to all in-neighbors; or else, (2)~if for some literal $(\ell, p(\ell))\in A(v)$, $k_i\in\psi(\ell)$ or $k_i\in\psi(p(\ell))$, then $v$ broadcasts $\langle\ell, 1\rangle$ to all in-neighbors; or else, (3)~if $v_i\neq\ ?$, $v$ broadcasts $\langle v_i, hop(v, v_i)\rangle$ to all in-neighbors; or else, (4)~for each in-neighbor $(u, p(u))\in \Gamma_{in}(v)$, if $k_i\in\psi(p(u))$, then $v$ sends $\langle v, 0\rangle$ to that in-neighbor $u$.

\section{Experimental Evaluation}\label{sec:results}
We now evaluate the performance of Quegel. The experiments were conducted on a cluster of 15 machines, each with 24 cores (two Intel Xeon E5-2620 CPU) and 48GB RAM. The machines are connected by Gigabit Ethernet. In Quegel, each worker corresponds to one process that uses a core. We ran 8 workers per machine (i.e., 120 workers in total) for all the experiments of Quegel, since running more workers per machine does not significantly improve query performance due to the limited network bandwidth. We used HDFS of Hadoop~1.2.1 on the cluster. The source code of Quegel and all algorithms discussed in this paper can be found in: \url{http://www.cse.cuhk.edu.hk/quegel}.

\begin{table}
\centering
\caption{Datasets (M = million)}\label{data}
\vspace{1mm}
\includegraphics[width=0.7\columnwidth]{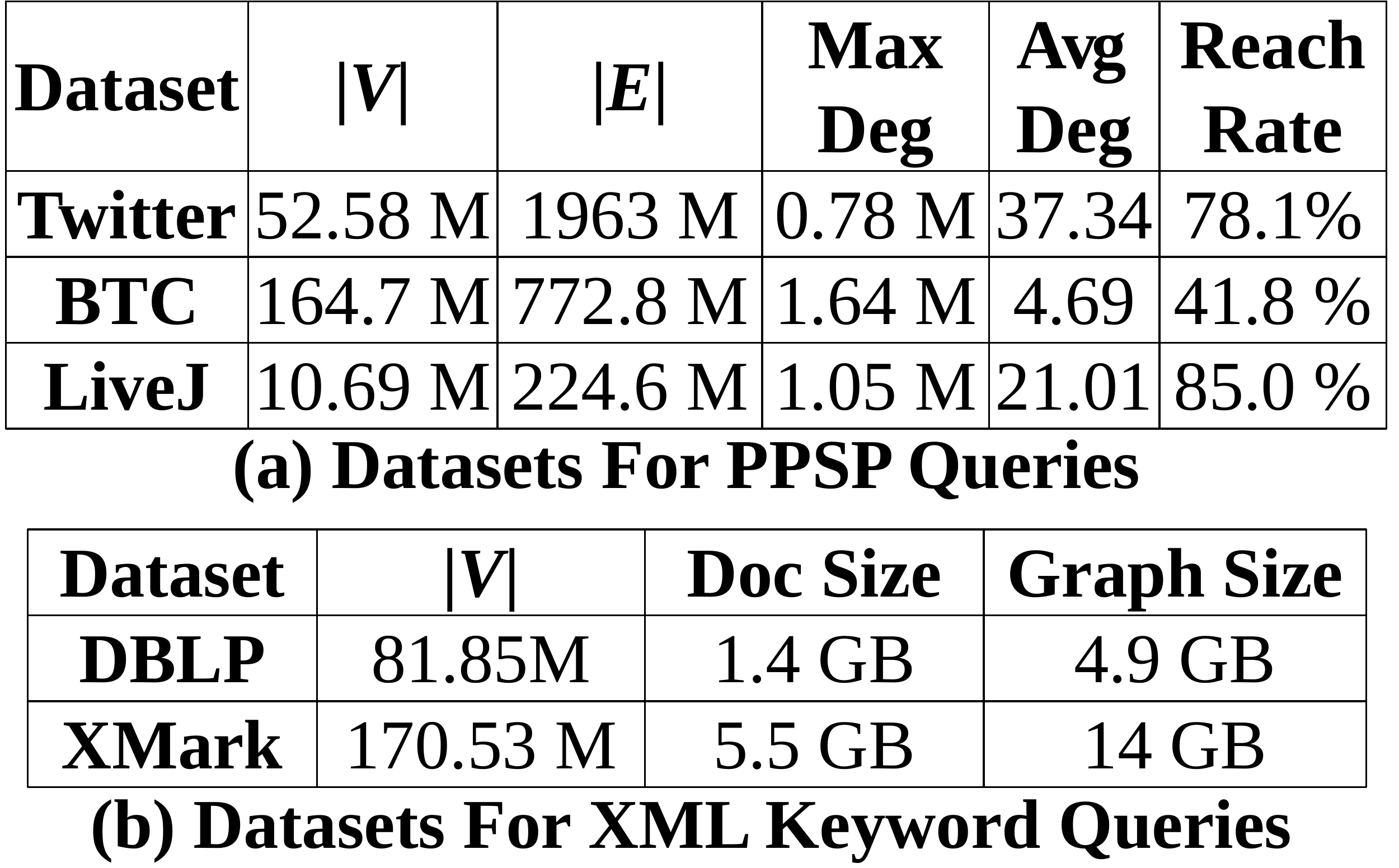}
\end{table}

\begin{table*}
\centering
\caption{Non-scalable systems: an illustration by answering 20 queries on {\em Livej} in serial}\label{neo4j}
\vspace{1mm}
\includegraphics[width=2.1\columnwidth]{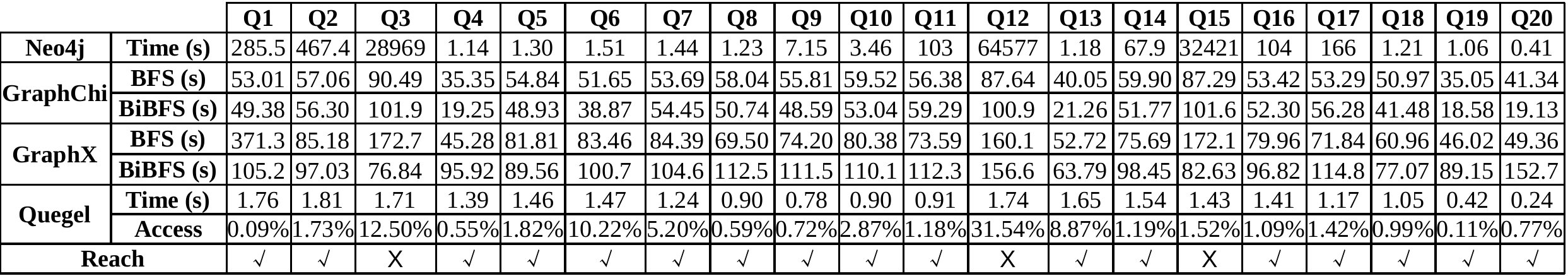}
\end{table*}

Table~\ref{data}(a) shows the datasets used in our experiments on PPSP queries: (i)~{\em Twitter}~\cite{Twitter}: Twitter who-follows-who network based on a snapshot taken in 2009; (ii)~{\em BTC}~\cite{BTC}: a semantic graph converted from the Billion Triple Challenge 2009 RDF dataset; and (iii)~a small dataset {\em LiveJ}~\cite{LiveJ} that refers to a bipartite network of LiveJournal users and their group memberships, which is used to demonstrate the poor scalability of some existing systems. {\em Twitter} is directed while {\em BTC} and {\em LiveJ} are undirected. Table~\ref{data}(a) also shows the maximum and average vertex degree of each graph, and we can see that the degree distribution is highly skewed. We randomly generate vertex pairs $(s, t)$ on each dataset for running PPSP query processing, and the percentage of queries where $s$ can reach $t$ is shown in column ``Reach Rate'' of Table~\ref{data}(a).

\vspace{1.6mm}

\noindent{\bf Comparison with Neo4j, GraphChi, and GraphX.} We first compared Quegel with a well-known graph database, Neo4j~\cite{neo4j}, a single-machine graph processing system, GraphChi~\cite{graphchi}, and the graph-parallel framework built on Spark (one of the most popular big data systems now), GraphX~\cite{graphx}. Neo4j and GraphChi were run on one of the machines in the cluster, while GraphX (shipped in Spark 1.4.1) was run on all the machines, using all cores and RAMs available.

The experimental results show that these three systems have poor scalability for processing PPSP queries. They either ran out of memory or are too time-consuming to process the two larger graphs, {\em Twitter} and {\em BTC}, and we were only able to record the results for them on the smallest dataset, {\em LiveJ}. Table~\ref{neo4j} reports the performance of these systems for answering 20 randomly-generated PPSP queries $(s, t)$ on {\em LiveJ}, where $s$ cannot reach $t$ in only three queries Q3, Q12, and~Q15.

Here, Quegel adopts the {\em Hub$^2$} algorithm described in Section~\ref{ssec:ppsp} and serves as a baseline to demonstrate why the other systems are not efficient. As preprocessing, it took 2912 seconds (end-to-end indexing time including graph loading/dumping) to compute the label set $L(v)$ for every vertex $v$ when 1000 hubs are selected. As Table~\ref{neo4j} shows, Quegel answers every query in around a second, and is able to support interactive querying. Row ``Access'' indicates the percentage of vertices accessed by Quegel, which is also small.

In contrast, Neo4j spent over 17 hours just to import {\em LiveJ}, and the imported graph consumes 64GB disk space while the dataset is only 1.1GB. In fact, {\em Twitter} could not be imported in several days while all disk space was used up when importing {\em BTC}. We used Neo4j's ``shortestPath'' function to answer the 20 PPSP queries, and as Table~\ref{neo4j} shows, the querying time is highly unstable and can be many hours if $s$ cannot reach $t$ because many vertices are visited.

The single-machine vertex-centric system GraphChi also cannot scale to process big graphs with reasonable query performance. As Table~\ref{neo4j} shows, GraphChi took tens of seconds to answer a query on the smallest graph  {\em LiveJ}, which is too slow for interactive querying. Moreover, BiBFS is not always faster than BFS since the additional field maintained (i.e., $d(v, t)$) increases data size, and BiBFS is much slower when $s$ cannot reach $t$ since the BFS from $t$ needs to reach all vertices in $t$'s connected component. Table~\ref{neo4j} also shows that GraphX is even slower than the single-machine system GraphChi. Moreover, GraphX could not process {\em BTC} and {\em Twitter} as it ran out of all RAM in our cluster. In fact, during the query processing on the smallest dataset {\em LiveJ}, GraphX already used more than half the RAM in each of our machines.

\vspace{1.6mm}

\noindent{\bf Comparison with Distributed Vertex-Centric Systems.} We compared Quegel with Giraph~1.0.0 and GraphLab~2.2 by running the BFS and BiBFS algorithms of Section~\ref{ssec:ppsp} for processing PPSP queries on the two large graphs, {\em Twitter} and {\em BTC}. For each dataset, we randomly generated 1000 queries $(s, t)$ for performance test. Since Giraph and GraphLab both use multi-threading, they have access to all the 24 cores on each of the 15 machines.

Giraph has a high start-up overhead since it needs to load the input graph $G$ from HDFS for the evaluation of each query; while GraphLab can keep $G$ in main memory for repeated querying after $G$ is loaded. Thus, we only ran Giraph for the first 20 queries. Table~\ref{q20twt} reports the cumulative time taken by Giraph, GraphLab and Quegel on {\em Twitter} for the first 20 queries. Note that the loading time of Giraph is contributed by all the 20 queries, while that of GraphLab and Quegel is the one-off loading time before query processing begins. Also, we specify Quegel to report query results on the console and hence there is no dumping time. For all experiments of Quegel, we set the capacity parameter $C=8$ unless otherwise stated. We also show the average access rate of a query (i.e., the percentage of vertices accessed) in the last row ``Access''.

\begin{table}
    \centering
    \caption{Cumulative time on {\em Twitter} (20 PPSP queries)}\label{q20twt}
    \vspace{1mm}
    \includegraphics[width=0.85\columnwidth]{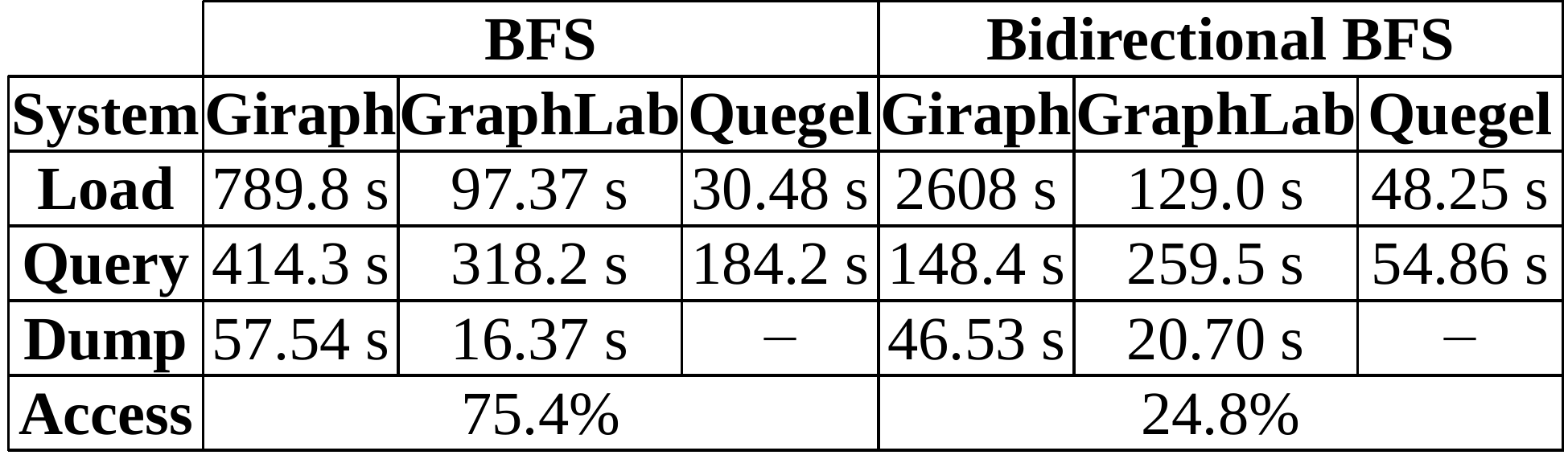}
\end{table}

\begin{table}
    \centering
    \caption{Cumulative time on {\em BTC} (20 PPSP queries)}\label{q20btc}
    \vspace{1mm}
    \includegraphics[width=0.85\columnwidth]{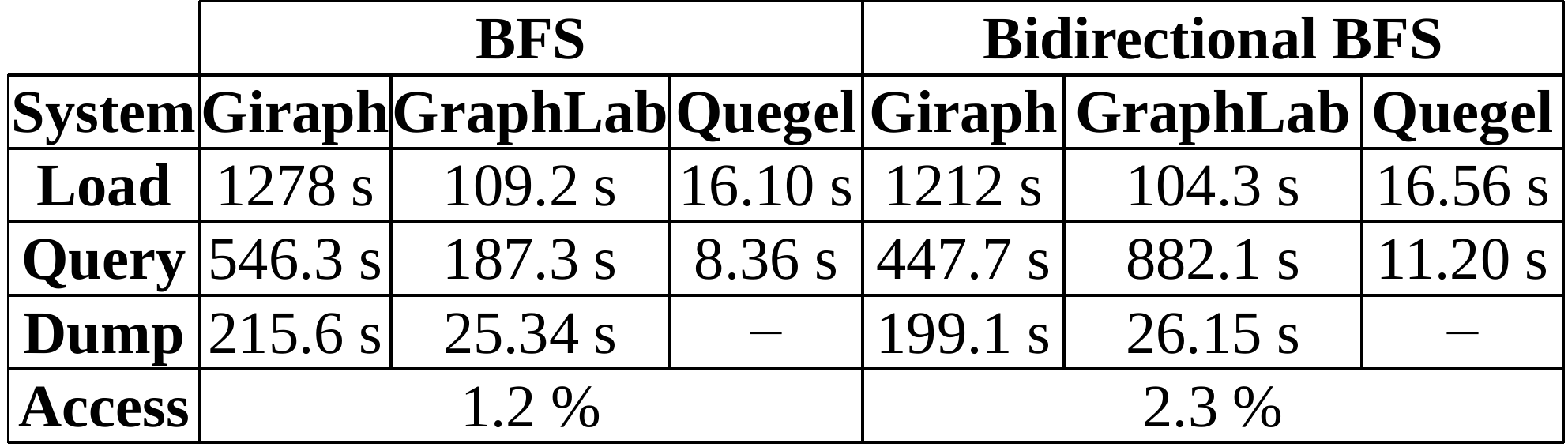}
\end{table}

\begin{table*}
    \begin{minipage}{0.55\linewidth}
        \centering
        \caption{Cumulative time on {\em Twitter} ($1k$ PPSP queries)}\label{q1000twt}
        \vspace{1mm}
        \includegraphics[width=\columnwidth]{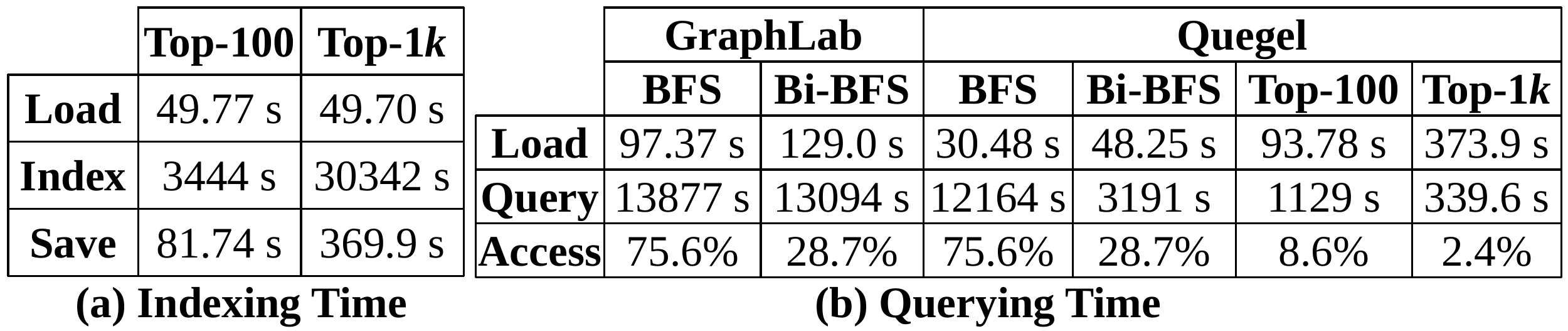}
    \end{minipage}
    \hfill
    \begin{minipage}{0.45\linewidth}
        \centering
        \caption{Cumulative time on {\em BTC} ($1k$ PPSP queries)}\label{q1000btc}
        \vspace{1mm}
        \includegraphics[width=\columnwidth]{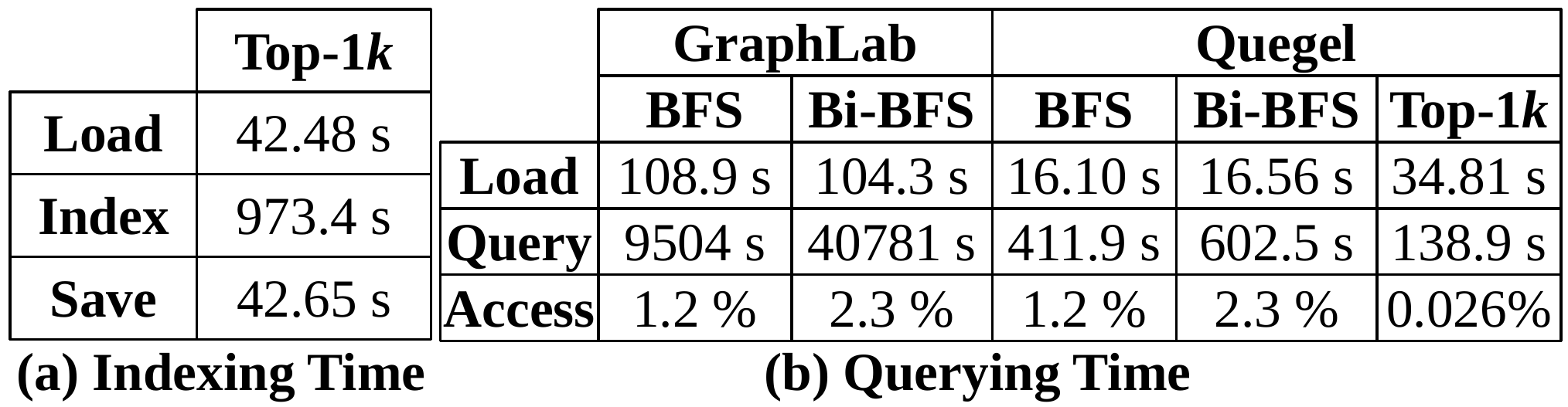}
    \end{minipage}
\end{table*}

Table~\ref{q20twt} shows that Quegel is significantly faster than Giraph and GraphLab when bidirectional BFS is used to process the queries. When BFS is used, Quegel is still considerably faster, but its computational time is 3 times longer than bidirectional BFS. This can be explained by the access rate of the queries, which actually demonstrates the effectiveness of Quegel's specialized design for querying workload that accesses only a small portion of the graph. The result also reveals that the other systems are not suitable for graph querying, as Giraph's loading time is already longer than its actual querying time, while GraphLab's querying time is unsatisfactory when the access rate is  small.

We remark that the loading time of Quegel for BiBFS is usually longer than that for BFS, since each vertex $v$ needs to load $\Gamma_{in}(v)$ in addition to $\Gamma_{out}(v)$. Note that $\Gamma_{in}(v)$ can be obtained by letting every vertex $u$ broadcast $\langle u\rangle$ to every out-neighbor $v\in\Gamma_{out}(u)$, and then let every vertex $v$ collect the IDs of its in-neighbors (i.e., $u$) to form $\Gamma_{in}(v)$ in the next superstep.

Table~\ref{q20btc} reports the performance of Giraph, GraphLab and Quegel on {\em BTC}. Since the access rate on {\em BTC} is much smaller than that on {\em Twitter}, the performance gap between Quegel and the other systems is significantly larger than on {\em Twitter}. This demonstrates that the design of Quegel is highly suitable for processing queries with small access rate. Another interesting observation is that, BFS has a smaller access rate than bidirectional BFS on {\em BTC}. This is because {\em BTC} consists of many connected components, and thus most queries $(s, t)$ are not reachable. For such a query, BFS terminates once all vertices in the connected component of $s$ are visited, while bidirectional BFS also visits the vertices in $t$'s connected component, leading to larger access rate.

\vspace{2mm}

\noindent{\bf Effect of Indexing.} We next show that using graph indexing, Quegel's query performance can be further improved by a large margin. Here, Quegel adopted the {\em Hub$^2$} the index.  For {\em Twitter}, we chose hubs as the top-$k$ vertices with the highest in-degree, out-degree, and their sum, and found that the results are similar. Thus, we only report the results for hubs with highest in-degree. Table~\ref{q1000twt}(a) shows the indexing time for $k=100$ and $k=1000$, which shows that each BFS from a hub takes about 30 seconds on average. Thus, $k$ cannot be too large in order to keep the preprocessing time reasonable.

Table~\ref{q1000twt}(b) shows the total time of processing all the 1000 queries using BFS and BiBFS in Quegel and GraphLab, and Quegel using the {\em Hub$^2$} index with $k=100$ and $k=1000$. We remark that Giraph is too expensive to process 1000 queries due to its high start-up overhead, and is thus not included. Clearly, when index is applied in Quegel, query performance is significantly improved, which can be explained by the reduction in the access rate. The result demonstrates the effectiveness of graph indexing. Although the loading time increases with larger $k$ since more core-hubs for each vertex are loaded from HDFS, this cost is not critical since graph loading is a one-off preprocessing operation and has no influence on subsequent query performance. When $k=1000$, {\em Hub$^2$} processes 1000 queries in 339.6 seconds, which is about 3 queries per second (about 39 times better than GraphLab).

We also build the {\em Hub$^2$} index on {\em BTC}, by picking the top-1000 vertices with the highest degree. Table~\ref{q1000btc}(a) shows the indexing time, which is very fast since {\em BTC} consists of many connected components and each BFS from a hub accesses only one component. Table~\ref{q1000btc}(b) shows the time of processing all the 1000 queries using BFS and BiBFS by Quegel and GraphLab, as well as by Quegel using the {\em Hub$^2$} index. We can see that the index significantly improves the query performance, and Quegel can process over 7 PPSP queries per second on {\em BTC} (68 times better than GraphLab).

\vspace{2mm}

\begin{table}
    \centering
    \caption{Effect of capacity and machine number}\label{q512twt}
    \vspace{1mm}
    \includegraphics[width=0.85\columnwidth]{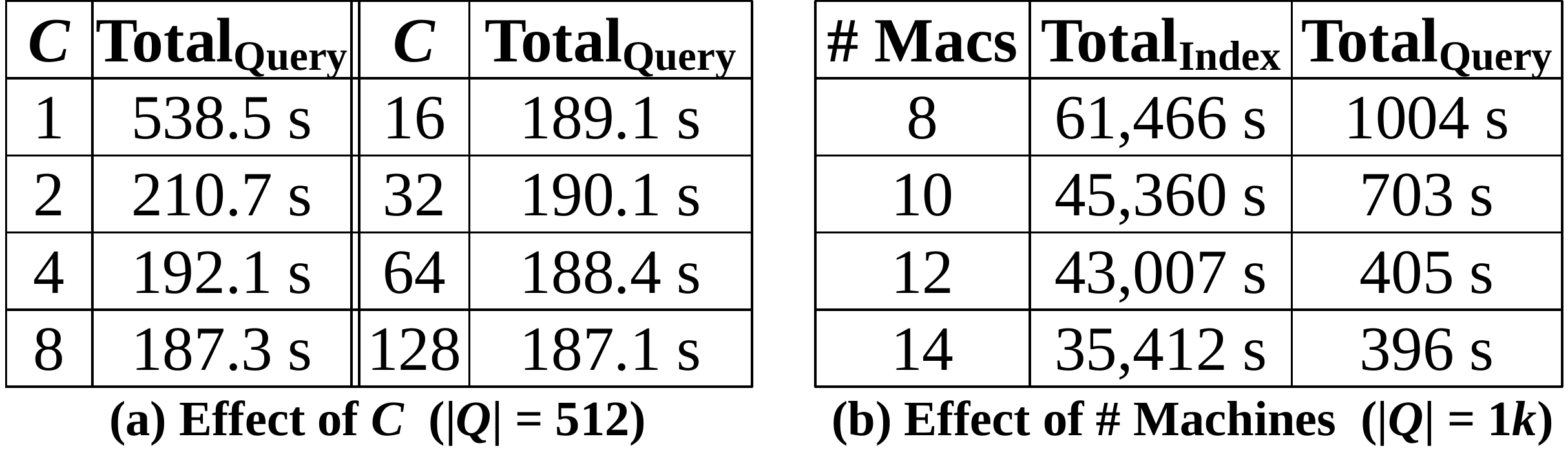}
\end{table}

\noindent{\bf Effect of Capacity Parameter.} We next examine how the capacity parameter $C$ influences the throughput of query processing in Quegel, by processing the first 512 queries with {\em Hub$^2$} ($k=1000$) on {\em Twitter} with different values of $C$. Table~\ref{q512twt}(a) reports the total time of processing the 512 queries, where we can see that processing queries one by one (i.e., $C=1$) is significantly slower than when a larger capacity is used. For example, when $C=8$, the total query processing time is only 1/3 of that when $C=1$, which verifies the effectiveness of superstep-sharing. However, the query throughput does not increase when we further increase $C$, since the cluster resources are already fully utilized. Similar results have also been observed on the other datasets.

Note that Quegel took 538.5 seconds to process 512 queries on {\em Twitter} when $C=1$, and thus the average time of processing a query is still around 1 second, which is similar to the response time on the small {\em Livej} dataset (see Table~\ref{neo4j}). This shows that the interactive querying performance of Quegel scales well to graph size.

\vspace{2mm}

\noindent{\bf Horizontal Scalability.} We now show how the performance of Quegel scales with the number of machines by processing the 1000 PPSP queries with {\em Hub$^2$} ($k=1000$) on {\em Twitter} with different cluster size. Table~\ref{q512twt}(b) reports the total time of indexing and processing the 1000 queries. The result shows that both the indexing time and querying time continue to decrease as the number of machine increases.

\begin{table}[t]
    \centering
    \caption{Results on XML keyword search}\label{xml_exp}
    \vspace{1mm}
    \includegraphics[width=0.8\columnwidth]{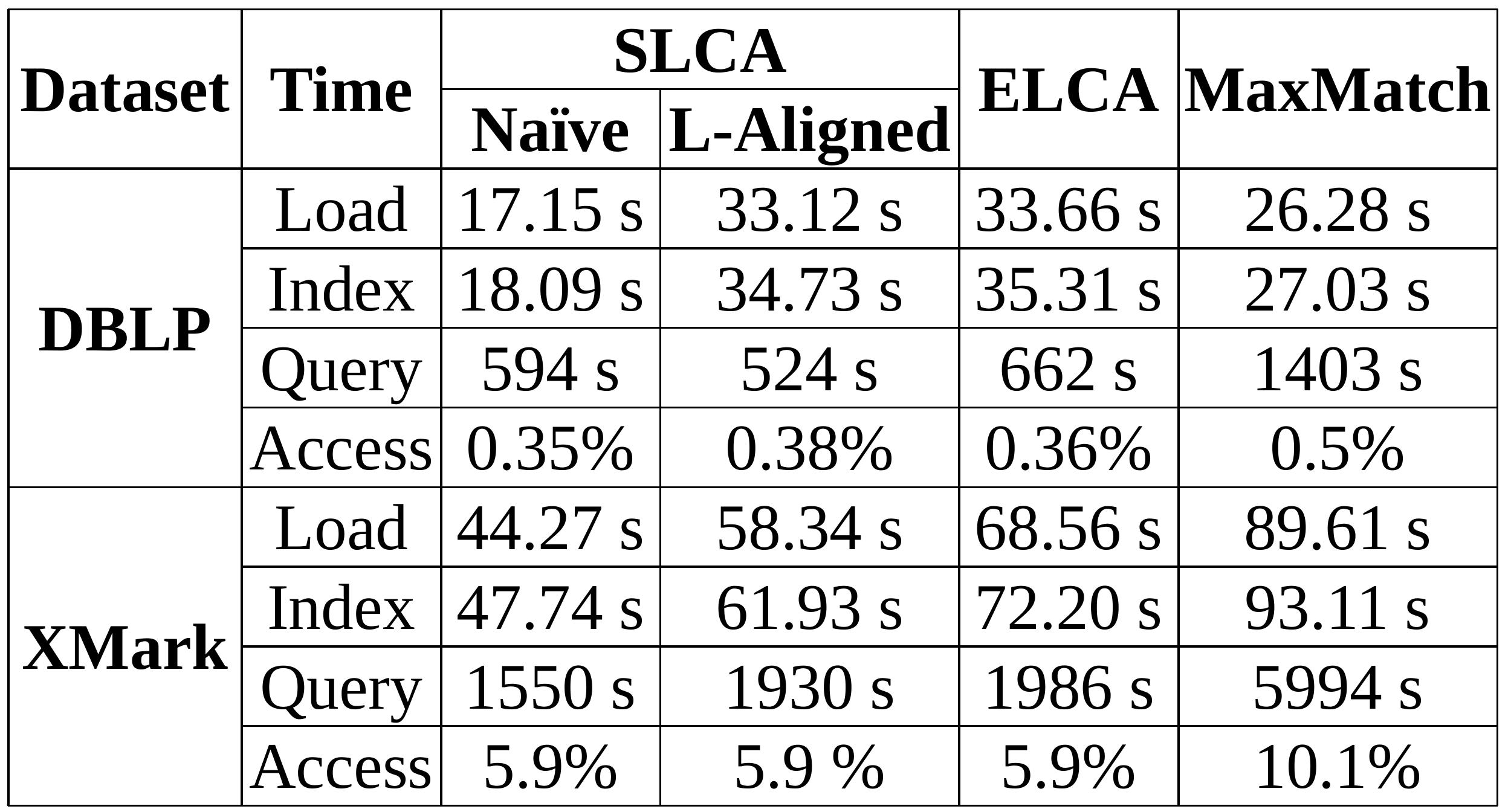}
\end{table}

\vspace{2mm}

\noindent{\bf Performance on XML Keyword Search.} We evaluate the performance of processing XML keyword queries in Quegel on the DBLP dataset~\cite{dblp}, as well as a larger XMark benchmark dataset~\cite{xmark}. Table~\ref{data}(b) shows the data statistics, where $|V|$ refers to the number of vertices in the XML tree, ``Doc Size'' refers to the file size of the raw XML document, and ``Graph Size'' refers to the file size of the parsed graph representation. Note that it is not clear how to implement our algorithms in other graph-parallel systems; for example, they do not provide a convenient way to construct and utilize inverted index. We generated 1000 keyword queries for each dataset, by randomly picking a query from a query pool each time, and repeating it for 1000 times. The query pools were obtained from existing work~\cite{TermehchyW11tods,ZhouBCL12dasfaa,ZhouBWLCLG12icde,ZhouB0ZM14vldb}, each contains tens of well-selected keyword queries.

Table~\ref{xml_exp} reports the performance of computing SLCA, ELCA and MaxMatch on the datasets using the 1000 queries. The reported metrics include the one-off graph loading and inverted index construction time, the total time of processing the 1000 queries (including result dumping), and the average access rate of a keyword query. The results verify that Quegel obtains good performance. For computing SLCAs, the average processing time of each query is only 0.5--0.6 second on DBLP. The time is longer on XMark, which is because XMark queries are less selective, and thus their access rates are much higher than those of DBLP queries. Finally, we can observe that the level-aligned version of SLCA algorithm is more efficient than the simple one on DBLP, but it is slower on XMark. This is because, vertices at the upper levels of DBLP's XML tree have high fan-outs, and thus the level-alignment technique significantly reduces the number of messages; while the vertex fan-out in XMark is generally small, and the cost incurred by the aggregator out-weighs the benefit of message reduction.

\begin{table}
\centering
\caption{Terrain Datasets (M = million)}\label{data_terrain}
\vspace{1mm}
\includegraphics[width=0.7\columnwidth]{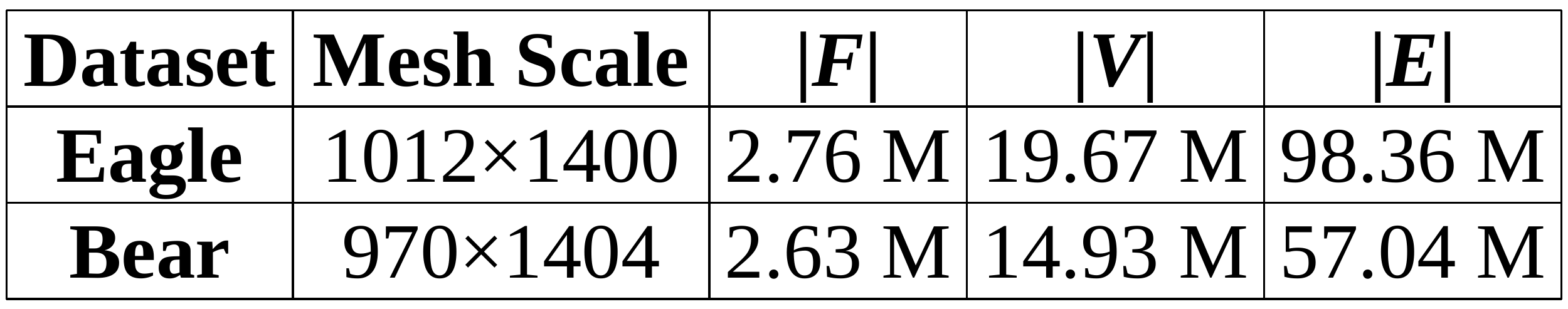}
\end{table}

\begin{table*}[t]
    \centering
    \caption{Results on terrain shortest-path queries}\label{terrain_exp}
    \vspace{1mm}
    \includegraphics[width=2\columnwidth]{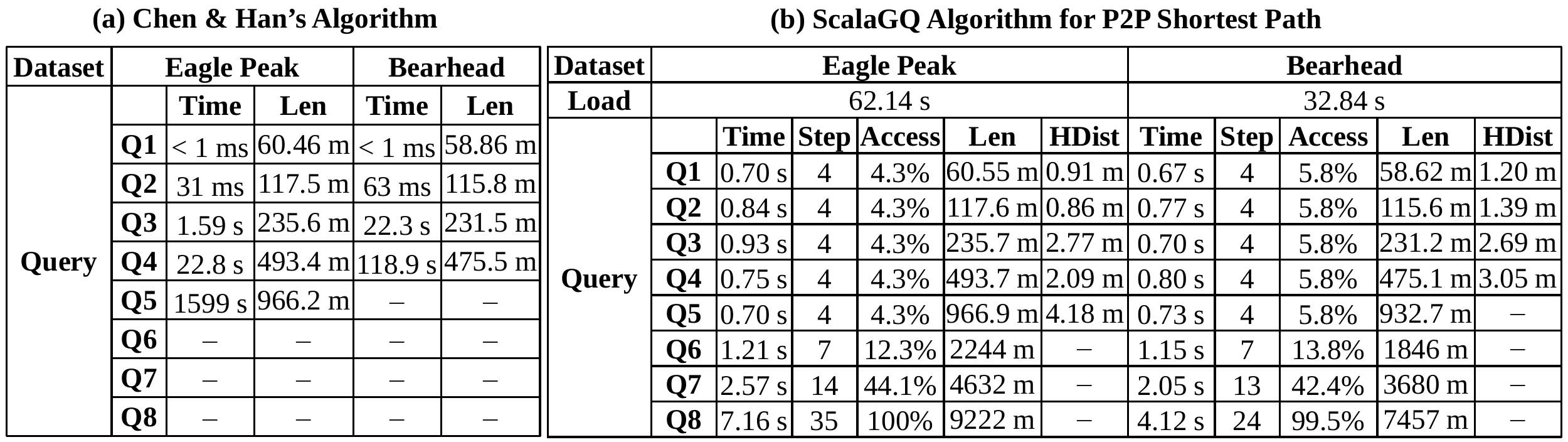}
\end{table*}

\vspace{2mm}

\noindent{\bf Experiments on terrain shortest-path queries.} We evaluate the performance of computing terrain shortest paths in Quegel using two real terrain datasets from USGS\footnote{\scriptsize http://data.geocomm.com/}, Eagle Peak ({\em Eagle}) and Bearhead ({\em Bear}). The data statistics are shown in Table~\ref{data_terrain}, and we explain those of {\em Eagle} below (similarly for {\em Bear}). The raw data of {\em Eagle} is a $1012\times1400$ elevation mesh of locations at 10m sampling interval; the TIN constructed from the mesh has $|F|=2.76$~M faces, while the network $G=(V, E)$ constructed using our approach with $\epsilon= 2$ meters has $|V|=19.67$~M and $|E|=98.36$~M. For each dataset, we pick a source location $s$ on the upper-left corner of the mesh, and then pick 8 destination points along the diagonal direction that are $2^2$, $2^3$, $\ldots$, $2^9$ cells away from $s$, to form 8 queries $Q1$, $Q2$, $\ldots$, $Q8$.

As a baseline, we evaluate the queries by running {\em Chen and Han's algorithm} (\emph{CH})~\cite{KanevaO00cccg} on the constructed TINs. Table~\ref{terrain_exp}(a) shows the computation time and shortest-path length for each query. However, when $s$ and $t$ are far from each other (e.g. $>1$ km), \emph{CH} ran out of memory, and the corresponding table entries are marked with ``--''. As Table~\ref{terrain_exp}(a) shows, when $t$ is close to $s$, \emph{CH} is very efficient. But as the distance increases, the computation time of \emph{CH} increases sharply. For example, it takes 1599 seconds to evaluate $Q5$ on Eagle Peak even though $t$ is only within 1 km from $s$.

We also process these queries in Quegel on the networks constructed by our approach, and the results are reported in Table~\ref{terrain_exp}(b). The results show that our Quegel algorithm scales to long paths. For example, it takes less than 1 second to compute a path that is less than 1km, and it takes only 7 seconds to compute a path over 9km. Moreover, when $s$ and $t$ are close to each other, our early termination technique effectively reduces the portion of network accessed.

As for the path quality, we compare a path $P_1$ computed by \emph{CH} with the path $P_2$ computed by Quegel for the same query. Comparing Table~\ref{terrain_exp}(a) with Table~\ref{terrain_exp}(b), we can see that the lengths of $P_1$ and $P_2$ are very close to each other. However, since they are computed from different terrain models (TIN v.s.\ our transformed network), it is not sufficient to compare only the path length. We show that the actual shapes of $P_1$ and $P_2$ are very similar as follows. Let $P$ and $P'$ be two polylines, $d(p, P)$ be the Euclidean distance from point $p$ to its closest point on $P$, and define $d(P, P')=\max_{p\in P}d(p, P')$. Then, the Hausdorff distance between $P_1$ and $P_2$ (which are 3D polylines) is given by {\em HDist}$(P_1, P_2)=\max\{d(P_1, P_2),\\
d(P_2, P_1)\}$~\cite{hangouet1995computation}, which captures how similar the shapes of $P_1$ and $P_2$ are. We compute {\em HDist}$(P_1, P_2)$ for all the queries (where $P_1$ is available), which are shown in column ``HDist'' in Table~\ref{terrain_exp}(b). The result is convincing; for example, for $Q5$ on {\em Eagle}, {\em HDist}$(P_1, P_2)$ is only 4 meters when the path lengths are around 966 meters.

\begin{figure}[t]
    \centering
    \includegraphics[width=0.8\columnwidth]{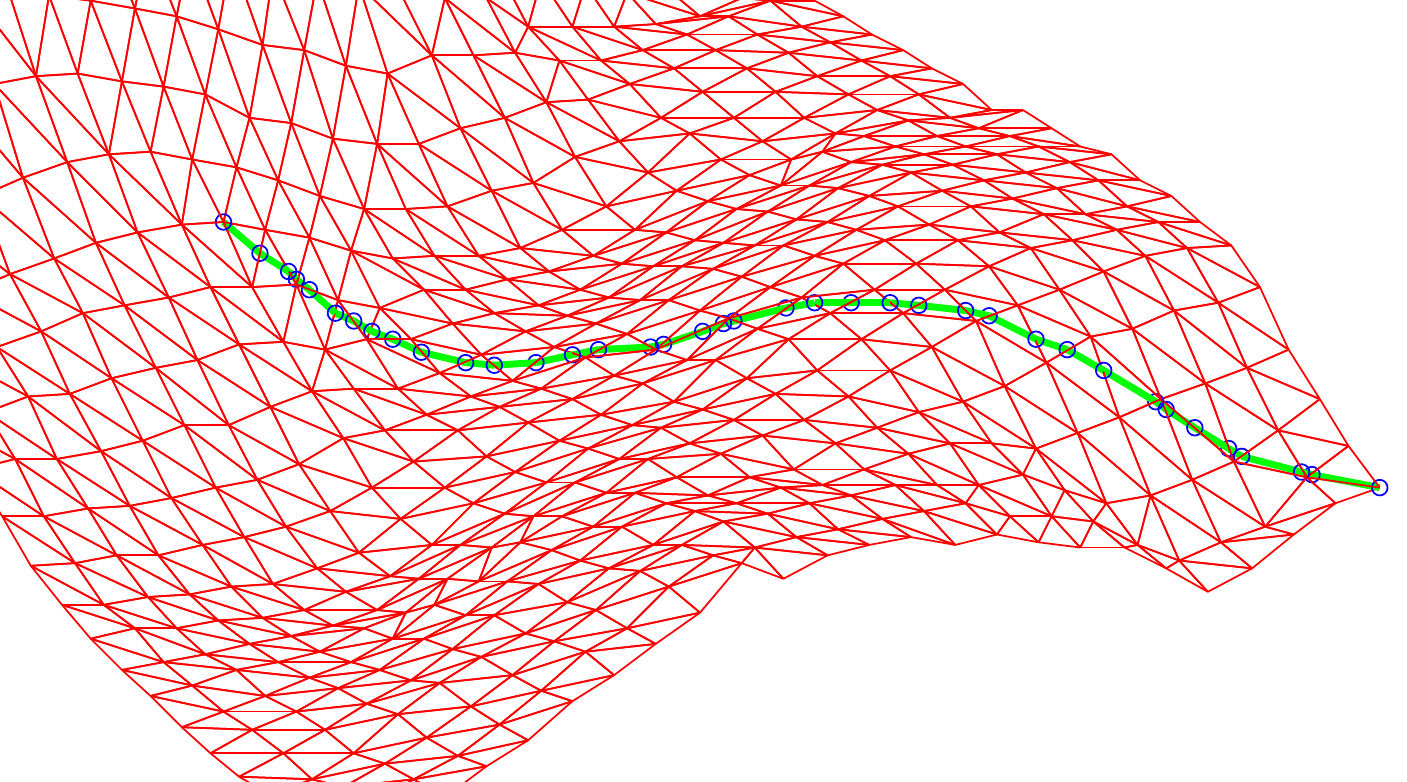}
    \caption{Path illustration (best viewed in color)}\label{paths}
\end{figure}

To visualize the path similarity, we plot $P_1$ and $P_2$ for query $Q3$ on {\em Eagle} in Figure~\ref{paths}, along with a fragment of the TIN. Since $P_1$ and $P_2$ are very similar, to show both paths clearly, we plot $P_1$ as a green line, while for $P_2$, we only show the points on this polyline using blue circles. As can be observed from Figure~\ref{paths}, the shapes almost coincide with each other. \vspace{2mm}

\vspace{2mm}

\begin{table}[!t]
\centering
\caption{Results on reachability queries}\label{reach_exp}
\vspace{1mm}
\includegraphics[width=0.7\columnwidth]{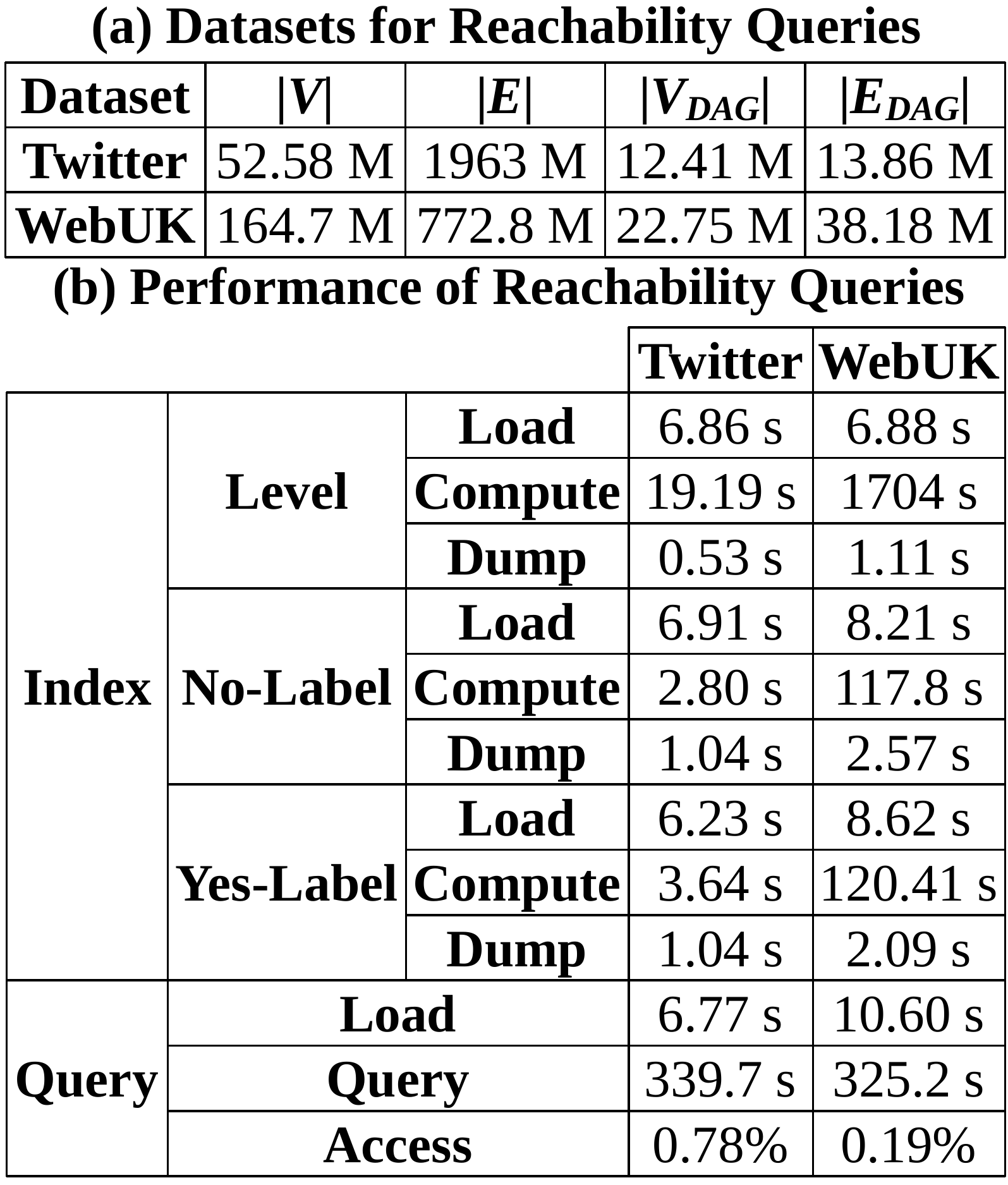}
\end{table}

\noindent{\bf Experiments on P2P reachability.} We now report the performance of processing P2P reachability queries in Quegel. The datasets used are shown in Table~\ref{reach_exp}(a): (1)~the {\em Twitter} dataset that was used in the experiments on PPSP queries; and (2)~{\em WebUK}\footnote{\scriptsize http://law.di.unimi.it/webdata/uk-union-2006-06-2007-05}: a web graph generated by combining twelve monthly snapshots of the .uk domain collected for the DELIS project. Table~\ref{reach_exp}(a) shows the number of vertices and edges of not only the original graph $G=(V, E)$, but also of the converted DAG $G_{DAG}=(G_{DAG}, E_{DAG})$.

Table~\ref{reach_exp}(b) shows the indexing performance of computing level labels, yes-labels and no-labels by a series of three Quegel graph-analytics jobs (Quegel also provides another kind of {\em Worker} class for programming Pregel-like tasks). We can see that computing level labels is relatively expensive, but once the level labels are available, the computation of yes-labels and no-labels is highly efficient. Jobs of {\em WebUK} take more time than those of {\em Twitter}, mainly due to more number of supersteps caused by the large diameter of {\em WebUK} (web graphs exhibit spatial locality and tend to have large diameter). For example, it takes Quegel 2793 supersteps to compute the level labels on {\em WebUK}, while the same task takes only 23 supersteps on {\em Twitter}.

To evaluate the querying performance of Quegel for P2P reachability queries, we randomly generate 1000 queries $(s, t)$, and the results are also shown in Table~\ref{reach_exp}(b), including the one-off graph loading time, the time of processing all the 1000 queries (results are dumped to HDFS), and the average access rate. We can see that it takes Quegel only 0.3 second on average to evaluate a P2P reachability query on such large graphs, which is very efficient.

\vspace{2mm}

\begin{table}[!t]
\centering
\caption{Results on RDF keyword search}\label{rdf_exp}
\vspace{1mm}
\includegraphics[width=0.75\columnwidth]{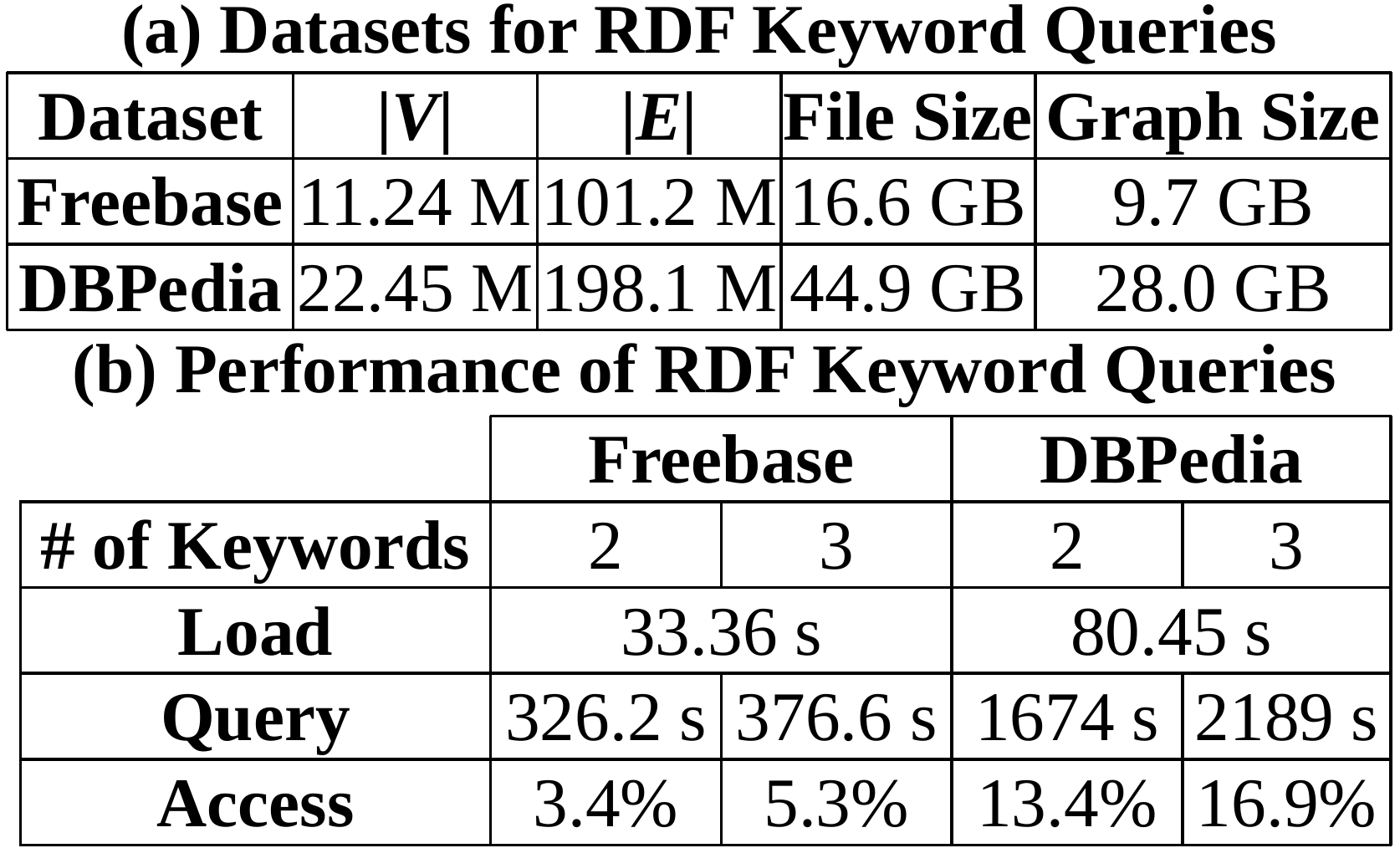}
\end{table}

\noindent{\bf Experiments on graph keyword search.} We now report the performance of processing graph keyword queries in Quegel. We use two RDF datasets from Billion Triples Challenge 2012\footnote{\scriptsize http://km.aifb.kit.edu/projects/btc-2012/}: {\em Freebase} and {\em DBPedia}, whose number of vertices (including resources and literals) and edges are shown in Table~\ref{rdf_exp}(a) as well as the RDF file size and the file size after converting to adjacency list representation.

We generate keyword queries on each dataset as follows. We select 30 most frequent words into a set $K_{30}$. Then, for each word $k\in K_{30}$, we activate every vertex $v$ with $k\in\psi(v)$ in Quegel, and run multi-source BFS for 3 hops to activate all vertices within 3 hops. We then collect the words contained in those vertices, and obtain the 100 most frequent words in predicates, denoted by $P_{100}(k)$, and the 100 most frequent words in non-predicates, denoted by $N_{100}(k)$. We then form 600 two-keyword queries $(k_1, k_2)$, where for each $k_1\in K_{30}$ we randomly select 20 words $k_2\in N_{100}(k_1)$; and form 600 three-keyword queries $(k_1, k_2, k_3)$, where for each $k_1\in K_{30}$, we randomly select 20 word pairs $(k_2, k_3)$ with $k_2\in P_{100}(k_1)$ and $k_3\in N_{100}(k_1)$. We generate queries in this way, so that $k_1$ has relatively low selectivity while $k_2$ and $k_3$ are also relevant to $k_1$.

Table~\ref{rdf_exp}(b) reports the performance of processing the graph keyword queries, including the one-off graph loading time, the total processing time of all 600 queries, and the average access rate. We can see that it takes Quegel only 0.3 second on average to evaluate a graph keyword query on {\em Freebase} and 1--2 seconds on average on {\em DBPedia}, which is very efficient given the big data size. Moreover, the querying time and access rate are higher when there are 3 keywords in the query, which shows that the query cost generally increases with the number of keywords in a query.

\section{Conclusions}   \label{sec:conclude}

We developed a distributed system, \textbf{Quegel}, for general-purpose querying of big graphs. To our knowledge, this is the first work that studies how to apply Pregel's user-friendly vertex-centric programming interface to efficiently process queries in big graphs. This is also the first general-purpose system that applies graph indexing to speed up query processing in a distributed platform. We demonstrated how Quegel is used to process five types of queries, i.e., PPSP queries, XML keyword queries, terrain shortest path queries, point-to-point  reachability  queries,  and  graph  keyword queries. We also showed Quegel obtained good performance for processing these queries.

For future work, we will continue to improve the performance of Quegel by designing various optimization techniques, proposing efficient graph indexes to be adopted in Quegel, and providing performance guarantees for answering different types of graph queries~\cite{ppa}. 

{\small
\bibliographystyle{abbrv}
\bibliography{ref_quegel,ref_full}
}

\end{document}